\documentclass[
  reprint,
  nofootinbib,
  amsmath,amssymb,
  aps,
  prd,
  floatfix,
  showkeys
]{revtex4-2}

\usepackage[USenglish]{babel}

\usepackage{amsmath}
\usepackage{amssymb}
\usepackage{amsfonts}
\usepackage{amsbsy}
\usepackage{bm}            
\usepackage{mathrsfs}
\usepackage{slashed}
\usepackage{extarrows}

\usepackage{graphicx}
\usepackage{graphics}
\usepackage{epstopdf}
\usepackage{dcolumn}       

\usepackage[dvipsnames]{xcolor}

\usepackage{array}
\usepackage{multirow}
\usepackage{enumitem}

\usepackage[caption=false]{subfig}

\usepackage{url}
\usepackage{varioref}
\usepackage{ulem}         
\usepackage{float}
\usepackage{verbatim}
\usepackage{latexsym}
\usepackage{orcidlink}
\usepackage{setspace}

\newcounter{widefn}

\newcommand{\LCDM}{$\Lambda$CDM}
\newcommand{\Om}{\Omega_{\mathrm{m},0}}
\newcommand{\Or}{\Omega_{\mathrm{r},0}}
\newcommand{\wa}{w_a}
\newcommand{\wo}{w_0}
\newcommand{\Dh}{\mathcal{D}_H}

\newcommand{\fsig}{f\sigma_8}
\newcommand{\cs}{c_s^2}
\newcommand{\skea}{\sqrt{-k}\,\eta_0}            
\newcommand{\code}[1]{\texttt{#1}}
\newcommand{\class}{{\sc class}}
\newcommand{\cobaya}{{\sc cobaya}\,}

\PassOptionsToPackage{
    colorlinks,
    citecolor=blue,
    urlcolor=magenta,
    linkcolor=blue
}{hyperref}
\usepackage{hyperref}

\begin{document}
\title{\bf General Relativistic Entropic Acceleration at the
perturbation level:\\ a {\sc class} implementation and first
Boltzmann-code constraints}

\author{Simone~D'Onofrio \orcidlink{0000-0001-7925-3905}}
\email{donofrio@ice.csic.es}
\affiliation{ Institute of Space Sciences (ICE, CSIC) C. Can Magrans s/n, 08193 Barcelona, Spain}

\author{Dong Ha Lee \orcidlink{0009-0000-2992-3157}}
\email{dhlee1@sheffield.ac.uk}
\affiliation{School of Mathematical and Physical Sciences, University of Sheffield, Hounsfield Road,
Sheffield S3 7RH, United Kingdom}

\author{Eleonora Di Valentino \orcidlink{0000-0001-8408-6961}}
\email{e.divalentino@sheffield.ac.uk}
\affiliation{School of Mathematical and Physical Sciences, University of Sheffield, Hounsfield Road,
Sheffield S3 7RH, United Kingdom}

\author{Juan Garc\'ia-Bellido \orcidlink{0000-0002-9370-8360}}
\email[]{juan.garciabellido@uam.es}
\affiliation{Instituto de F\'isica Te\'orica UAM-CSIC, Universidad Auton\'oma de Madrid, Cantoblanco 28049 Madrid, Spain}

\date{\today}

\begin{abstract}
General Relativistic Entropic Acceleration (GREA) attributes the late-time
acceleration of the Universe to the entropy growth of the causal cosmological
horizon, without a cosmological constant, with a phenomenology fixed by the
single $\mathcal{O}(1)$ parameter $\alpha$. The model has so far been confronted
with data only at the background level. We present its first implementation
within an Einstein--Boltzmann solver: the GREA background is integrated directly
into \class, while the entropic component is evolved as an effective fluid
regulated by the parametrized-post-Friedmann scheme, giving access to the full
CMB and matter power spectra. A Markov-chain Monte Carlo analysis with \cobaya
against the full primary-CMB likelihoods, DESI DR2 BAO and Type Ia supernovae
constrains the coupling $\alpha \sim 1$, in excellent agreement
with the theoretical prediction, with a fit matching
$\Lambda$CDM to within $|\Delta\chi^2| \lesssim 6$ despite the addition of a single free
parameter. The equation of state inferred from the data agrees with binned,
model-independent reconstructions and exhibits a second crossing of the phantom divide
at $z \simeq 2$, a distinctive prediction of the thermodynamic dynamics rather
than of an imposed parametrization.
\end{abstract}

\maketitle

\section{Introduction}
\label{sec:intro}

The origin of the late-time accelerated expansion of the Universe remains
one of the deepest open problems in physics. Ever since it was inferred
from the dimming of distant Type~Ia supernovae~\citep{SupernovaSearchTeam:1998fmf,SupernovaCosmologyProject:1998vns},
the standard cosmological model (\LCDM) has accounted for the acceleration
through a cosmological constant $\Lambda$~\citep{Sahni:1999gb}. This model
provides a remarkable fit to a broad range of observations, from the cosmic
microwave background~\citep{Planck:2018vyg,Planck:2018nkj,ACT:2025fju,SPT-3G:2025bzu} to the clustering
of galaxies and the baryon acoustic oscillation (BAO) feature~\citep{eBOSS:2020yzd,eBOSS:2021pff,DESI:2025fii}.
Its very success, however, throws its shortcomings into sharper relief: the
measured value of $\Lambda$ lies many orders of magnitude below any natural
quantum-field-theory expectation, giving rise to the celebrated cosmological-constant and
coincidence problems~\citep{Weinberg:1988cp,Carroll:2000fy}, and the model
offers no fundamental account of either of its dark components.

These theoretical concerns are compounded by a growing set of observational
tensions and anomalies~\citep{Bull:2015stt,Bullock:2017xww,Perivolaropoulos:2021jda,Abdalla:2022yfr,Calderon:2023obf,CosmoVerseNetwork:2025alb}.
The most statistically significant is the discrepancy between the local and
the CMB-inferred value of the Hubble constant~\citep{Riess:2021jrx,DiValentino:2021izs,Verde:2019ivm,DiValentino:2020zio,Schoneberg:2021qvd,Shah:2021onj,DiValentino:2022fjm,Kamionkowski:2022pkx,Giare:2023xoc,Hu:2023jqc,Verde:2023lmm,DiValentino:2024yew,Ong:2025cwv,Cai:2026swf},
while a milder disagreement persists in the amplitude of matter
fluctuations, $S_8$, probed by weak lensing and redshift-space
distortions~\citep{HSC:2018mrq,DiValentino:2020vvd,DiValentino:2020vvd,DiValentino:2018gcu,Nunes:2021ipq,DES:2021bvc,DES:2021vln,KiDS:2020suj,Asgari:2019fkq,Joudaki:2019pmv,DAmico:2019fhj,Kilo-DegreeSurvey:2023gfr,Troster:2019ean,Heymans:2020gsg,Dalal:2023olq,Chen:2024vvk,ACT:2024okh,DES:2024oud,Harnois-Deraps:2024ucb,Dvornik:2022xap,DES:2021wwk,Wright:2025xka,DES:2026fyc,DES:2026mkc}. More recently, the Dark Energy
Spectroscopic Instrument (DESI), combined with supernovae and CMB data, has
reported hints at the several-$\sigma$ level of a departure from a
$\Lambda$CDM expansion history~\citep{DESI:2024mwx,DESI:2025fii,DESI:2025zgx}.
If they hold up, such results would constitute the first evidence for new
physics beyond \LCDM. They are most often interpreted as a signature of
dynamical dark energy~\citep{DESI:2024aqx,Ishak:2024jhs,DESI:2025wyn,DESI:2024kob,Cortes:2024lgw,Shlivko:2024llw,Luongo:2024fww,Gialamas:2024lyw,Wang:2024dka,Ye:2024ywg,Tada:2024znt,Carloni:2024zpl,Chan-GyungPark:2024mlx,Bhattacharya:2024hep,Reboucas:2024smm,Najafi:2024qzm,Giare:2024gpk,Giare:2024oil,Jiang:2024xnu,RoyChoudhury:2024wri,Giare:2024oil,Giare:2025pzu,Kessler:2025kju,RoyChoudhury:2025dhe,Scherer:2025esj,Wolf:2025jlc,Santos:2025wiv,Specogna:2025guo,Cheng:2025lod,Cheng:2025hug,Ozulker:2025ehg,Li:2025vuh,Lee:2025pzo,Fazzari:2025lzd,Smith:2025icl,Herold:2025hkb,Cheng:2025yue,Gokcen:2026pkq,Ishak:2025cay,Najafi:2026kxs,Yang:2026yaq,Kessler:2026dbi,Lee:2026yzs,Li:2026asg,Giare:2026oti,GuptaChoudhury:2026gsl},
but alternative interpretations abound, invoking non-standard neutrino
sectors~\citep{Elbers:2024sha,Elbers:2025vlz,Craig:2024tky,Green:2024xbb,Elbers:2025vlz,Graham:2025dqn,Pulido-Hernandez:2026hcs,Yang:2026yaq,Loverde:2024nfi,Kibris:2026cqq}, an evolving dark-matter
component~\citep{Chen:2025wwn}, or early-time and geometric
solutions~\citep{DESI:2025ffm,Chaussidon:2025npr,Chen:2025mlf}.

Confronted with these hints, most phenomenological studies fall back on
arbitrary parametrizations of the dark-energy equation of state, $w(z)$,
the $w_0w_a$CDM form~\citep{Chevallier:2000qy,Linder:2002et,Caldwell:2005tm,dePutter:2008wt}
being the most widespread, or on model-independent reconstruction
techniques~\citep{Shafieloo:2005nd,Holsclaw:2010nb,Shafieloo:2012ht,Nesseris:2012tt,Calderon:2022cfj,Calderon:2022cfj}.
Such approaches offer flexibility but limited physical insight, and they
risk biasing the inference when the true $w(z)$ departs from the assumed
functional form. An intermediate route regularizes the reconstruction with a theory-informed prior, for instance one built on the general Horndeski class of scalar-tensor theories, retaining much of this flexibility while injecting physical structure into the inference \cite{Pogosian:2021mcs}. A still more informative strategy is to test models grounded in fundamental principles, whose acceleration is predicted rather than fitted.

General Relativistic Entropic Acceleration (GREA) is one such
framework~\citep{Garcia-Bellido:2021idr,Espinosa-Portales:2021cac}. It
provides a covariant formulation of out-of-equilibrium thermodynamics in
general relativity, in which the explicit breaking of time-reversal
invariance through entropy production drives an entropic force that behaves like
a bulk viscosity with negative effective pressure. When the entropy growth
is associated with a causal horizon, its thermodynamic effect is captured by
the Gibbons--Hawking--York boundary term~\citep{Gibbons:1976ue,York:1972sj},
so that the source of spacetime curvature becomes the Helmholtz free energy
$F=U-TS$ rather than the energy density alone. In the cosmological setting,
the present acceleration then arises from the growing entropy of the
cosmic~\citep{Garcia-Bellido:2021idr} and black-hole~\citep{Garcia-Bellido:2024tip}
horizons, without any cosmological constant. Remarkably, this thermodynamic
picture admits a holographic interpretation in which a bulk observer performing
long-range electromagnetic and gravitational measurements cannot
distinguish the acceleration induced by $\Lambda$ from that induced by the
horizon's boundary degrees of freedom~\citep{Garcia-Bellido:2025hji}. The
resulting phenomenology is fixed by a single $\mathcal{O}(1)$ parameter,
$\alpha$, the ratio of the spatial-curvature scale to the causal horizon
today: a given $\alpha$ determines a unique expansion history and definite
predictions for every background observable, addressing the coincidence
problem and, by shifting the matter-to-acceleration transition to higher
redshift, easing the Hubble tension~\citep{Garcia-Bellido:2024qau}.

GREA has so far been confronted with data only at the background level.
Already before the recent DESI and DES-SN5YR~\citep{DES:2024jxu,DESI:2024jis}
results, it was found to describe the cosmological observations as well as
\LCDM~\citep{Arjona:2021uxs}, and the detailed background and linear-growth
predictions from the homogeneous cosmic horizon were worked
out in~\citep{Garcia-Bellido:2024qau}. The most complete analysis to
date~\citep{Calderon:2025dhj}, using DESI~DR2 BAO together with several
supernova compilations and compressed CMB distance priors, finds a fit comparable
to \LCDM\ with $\alpha\simeq1.09$ and a transient phantom crossing at
$z\lesssim2$, while an independent study~\citep{Graziotti:2026qzl}
reproduces these constraints and forecasts future surveys. In both, the
goodness of fit is comparable to \LCDM, yet the Bayesian model comparison
tends to prefer the cosmological constant once the compressed CMB distance
priors are included~\citep{Graziotti:2026qzl}. A limitation
shared by all of this work, and flagged by the authors as the natural next
step, is that the cosmic microwave background enters only through these compressed
distance priors, and the perturbation sector is treated at most
semi-analytically, with no power spectrum or lensing spectrum ever computed
inside a Boltzmann code.

The present work removes that limitation. We report the first
implementation of GREA inside an Einstein--Boltzmann solver: the GREA
background is integrated directly in \class, while the entropic component is
described at the linear level as an effective dark-energy fluid with a
prescribed equation of state $w(a)$ and a bracketed sound speed, regulated
by the parametrized-post-Friedmann scheme so that the perturbations remain
regular across the transient phantom crossing. This gives access, for the
first time, to the full CMB temperature, polarization, and lensing spectra,
the linear matter power spectrum, and $\fsig(z)$ of the model, and it allows us
to perform a Markov-chain Monte Carlo analysis against the full primary-CMB
likelihoods rather than compressed distance priors. The paper is organized as
follows. Section~\ref{sec:framework} reviews the GREA framework and its
background dynamics; Section~\ref{sec:perturbations} sets out the linear
perturbations and the effective-fluid description; Section~\ref{sec:implementation}
details the \class\ implementation and its validation;
Section~\ref{sec:data} describes the data and methodology; and
Sections~\ref{sec:results} and~\ref{sec:conclusions} present our results and
conclusions.

\section{The GREA framework}
\label{sec:framework}

\subsection{Horizons and non-equilibrium thermodynamics}

GREA rests on the covariant formulation of non-equilibrium thermodynamics
in general relativity developed in
\citep{Garcia-Bellido:2021idr,Espinosa-Portales:2021cac}.
Incorporating the first law of thermodynamics into Einstein's equations in
the presence of entropy production extends the field equations with an
entropic-force tensor $f_{\mu\nu}$,
\begin{equation}
  R_{\mu\nu}-\tfrac{1}{2}R\,g_{\mu\nu}
   = \frac{8\pi G}{c^4}\,\big(T_{\mu\nu}-f_{\mu\nu}\big),
  \label{eq:einstein}
\end{equation}
where $f_{\mu\nu}$ encodes a bulk-viscosity-like contribution with negative
effective pressure $p_S=-T\,\mathrm{d}S/\mathrm{d}V<0$ dictated by the
second law~\citep{Arjona:2021uxs,Garcia-Bellido:2024qau}. The energy
conservation equation correspondingly acquires a source,
\begin{equation}
  \dot{\rho}+3H(\rho+p)=\frac{T\dot{S}}{a^{3}},
  \label{eq:continuity}
\end{equation}
where $S$ is the entropy per comoving volume and $T$ is the horizon temperature.
The construction follows from extending the Einstein--Hilbert action by the
Gibbons--Hawking--York boundary term~\citep{Gibbons:1976ue}, which allows one
to assign a temperature $T_H$ and an entropy $S_H$ to the causal horizon
$d_H=a\eta$ (with $\eta$ the conformal time),
\begin{equation}
  k_B T_H=\frac{\hbar}{2\pi}\,
           \frac{a\,\sinh\!\big(2\eta\sqrt{-k}\big)}{d_H^{2}\sqrt{-k}},
  \qquad
  S_H=\frac{k_B\,\pi}{\hbar}\,\frac{d_H^{2}}{G},
  \label{eq:TS}
\end{equation}
so that the associated energy density is $\rho_H=T_HS_H/a^{3}$. The factor
$\hbar$ signals the quantum-gravitational origin of the horizon
thermodynamics, used here as an effective description on a classical
background~\citep{Garcia-Bellido:2025hji}. The acceleration is driven by the
breaking of time-reversal invariance through entropy production; it is an
out-of-equilibrium effect rather than vacuum energy, in the lineage of
the entropic and emergent-gravity program~\citep{Jacobson:1995ab}.

\subsection{Background dynamics and the equation of state}

GREA is naturally formulated in an open ($k<0$) universe. Writing the
rescaled conformal time $\tau\equiv H_0\eta=H_0\!\int\!\mathrm{d}t/a$, the
first Friedmann equation takes the
form~\citep{Garcia-Bellido:2024qau,Calderon:2025dhj}
\begin{align}
  \tau'\equiv&\frac{\mathrm{d}\tau}{\mathrm{d}a}\\
   =&\left[\,a^{2}\sqrt{\;\Om\,a^{-3}\Big(1+\tfrac{a_{\rm eq}}{a}\Big)
      +\frac{4\pi}{3a^{2}}\,
       \frac{\sinh(2\tau)}{(-k)^{3/2}V_c}}\;\right]^{-1},
  \label{eq:background}
\end{align}
where $a_{\rm eq}=\Or/\Om$ and the comoving volume is
\begin{equation}
    (-k)^{3/2}V_c=\pi\big[\sinh(2\sqrt{-k}\,\eta_0)-2\sqrt{-k}\,\eta_0\big].
\end{equation}
The second term under the square root plays the role of dark energy. Its
$\sinh(2\tau)$ dependence is exponentially negligible during matter
domination and grows only at late times, which addresses the coincidence
problem. Given the dimensionless particle horizon
\begin{equation}
      \Dh(a)\equiv H(a)\,d_H(a)=Ha\,\eta,
\end{equation}
the phenomenology is set by a single $\mathcal{O}(1)$ parameter, $\alpha$, such that
\begin{equation}
  \alpha\,\Dh(z=0)=\skea,  \label{eq:alpha}
\end{equation}
the ratio of the curvature scale to the causal horizon today; a given
$\alpha$ fixes a unique expansion history, with best-fit values clustering
around $\alpha\simeq1$~\citep{Calderon:2025dhj}. One subtlety matters for
any code implementation: the present rate $H(z{=}0)$ is a derived quantity,
obtained by integrating \eqref{eq:background} up to $a=1$, while the $H_0$
in the definition of $\tau$ is only a normalization, so in general
$H(z{=}0)\neq H_0$~\citep{Garcia-Bellido:2024qau,Graziotti:2026qzl}. We
return to this in Section~\ref{sec:implementation}.

Defining an effective dark-energy density through
$\Omega_{\rm DE,0}\,f_{\rm DE}(z)\equiv E^2(z)-\Om(1+z)^3-\Or(1+z)^4$, with
$E\equiv H/H(z=0)$, the GREA equation of state follows in closed
form~\citep{Calderon:2025dhj},
\begin{align}
  w(a)=&-\frac{1}{3}\frac{\mathrm{d}\ln f_{\rm DE}}{\mathrm{d}\ln a}-1 \label{eq:f_def}\\
      =&-\frac{1}{3}\Big[\,2\,a\,\tau'(a)\,\coth\!\big(2\tau(a)\big)+1\,\Big].
  \label{eq:weff}
\end{align}
For $\alpha\simeq1$ this gives a transient phantom crossing at
$z\lesssim2$ (Fig.~\ref{fig:w(z)}), with a present slope
$\wa\equiv\mathrm{d}w/\mathrm{d}a|_{a=1}\simeq-0.3$ and a value
$\wo\equiv w(z{=}0)\simeq-1$~\citep{Calderon:2025dhj}. The function is
curved in $a$, so the Chevallier--Polarski--Linder
pair~\citep{Chevallier:2000qy,Linder:2002et} $(\wo,\wa)$ captures only its
very-low-$z$ tangent, and GREA is not equivalent to $w_0w_a$CDM over the
redshift range probed by the data. By shifting the matter-to-acceleration
transition to higher redshift, GREA raises $H(z{=}0)$ relative to \LCDM\ at
fixed early-time physics, which is how it addresses the Hubble
tension~\citep{Garcia-Bellido:2024qau,DiValentino:2021izs}; the
acceleration is transient, with the entropic term eventually diluting like
matter so that the Universe tends to a Minkowski rather than a de~Sitter
state. Two extensions of the picture are an inhomogeneous source, in which
the entropy growth of accreting supermassive black holes contributes to
the acceleration~\citep{Garcia-Bellido:2024tip}, and a holographic dual in
which a bulk observer cannot distinguish the entropic acceleration from
$\Lambda$ at the background level~\citep{Garcia-Bellido:2025hji}.

GREA has so far been compared with data only at the background level. The
earliest analysis found it to describe SNe~Ia and BAO comparably to
\LCDM~\citep{Arjona:2021uxs}. The most complete study to
date~\citep{Calderon:2025dhj} uses DESI~DR2 BAO~\citep{DESI:2025zgx}, three
SN~Ia compilations (Pantheon+~\citep{Brout:2022vxf},
Union3~\citep{Rubin:2023ovl}, DES-SN5YR~\citep{DES:2024jxu}) and compressed
CMB distance priors, finding a fit comparable to \LCDM\ with $\alpha\simeq1.09$,
while a phenomenological $w_0w_a$CDM model closer to the model-independent
reconstructions~\citep{DESI:2025fii} does better than both. An independent
analysis~\citep{Graziotti:2026qzl} reproduced these constraints and
forecasts future surveys, with a Bayesian preference for \LCDM\ once
compressed CMB data are added. Two limitations are shared by all of this
work and flagged by the authors as future tasks: the CMB enters only as
compressed distance priors, never as a full likelihood, and the perturbation
sector is used at most semi-analytically, with no power spectrum or
angular/lensing spectrum computed inside a Boltzmann code. The present work
removes both.

\section{Linear perturbations and the effective-fluid description}
\label{sec:perturbations}

At the linear level, Ref.~\citep{Garcia-Bellido:2024qau} derives the growth
of matter fluctuations on a homogeneous GREA background. The density
contrast $\delta\equiv\delta\rho_m/\bar\rho_m$ obeys
\begin{equation}
  a(\tau)\,\delta''(\tau)+a'(\tau)\,\delta'(\tau)
   =\tfrac{3}{2}\,\Om\,\delta(\tau),
  \label{eq:growth}
\end{equation}
with primes denoting derivatives with respect to conformal time. The source
is the standard $\tfrac{3}{2}\Om\,\delta$: GREA introduces no dark-energy
clustering term and no modification of the effective gravitational
coupling, and enters only through the modified scale-factor evolution
$a(\tau)$ in the Hubble-friction term. The sub-horizon growth, and with it
$D(a)$, $f(z)$, $\fsig(z)$, and the linear $P(k)$ shape, is therefore set
entirely by the GREA background, as for any smooth dark-energy model within
general relativity. Reproducing the semi-analytic $\fsig(z)$ of
\citep{Garcia-Bellido:2024qau} is consequently a validation target
rather than a new result.

The direction of the growth modification deserves a comment. Since GREA
prolongs the acceleration epoch, one might expect suppressed late-time
growth and some relief of the $\sigma_8/S_8$ tension. Solving
\eqref{eq:growth} on the validated background gives the opposite: at fixed
primordial amplitude $A_s$, GREA enhances the growth and yields a larger
$\sigma_8$ than \LCDM\ (Fig.~\ref{fig:growth}), in agreement with Fig.~8 of
Ref.~\citep{Garcia-Bellido:2024qau}. As a forward prediction, GREA therefore
does not relieve the $S_8$ tension, contrary to what the background
behavior alone might suggest.

A fully consistent Einstein--Boltzmann treatment would require the
perturbed entropic sector, namely $\delta f_{\mu\nu}$ and the resulting
fluid equations for $(\delta_{\rm GREA},\theta_{\rm GREA})$ with their sound
speed and anisotropic stress. The difficulty is concentrated in
$\delta f_{\mu\nu}$: the bulk-viscosity coefficient scales as
$\zeta\propto T\dot S\propto d_H^{2}=(a\eta)^{2}$, so its perturbation
involves $\delta d_H$, the perturbation of a light-cone integral and hence
a non-local quantity. Whether this non-local piece can be neglected
($\delta\zeta=0$) is the central open question of a complete derivation. We
do not settle it here, and defer the derivation of $\delta f_{\mu\nu}$,
feasible with \code{xAct} starting from the bulk-viscosity formalism of
Gagnon \& Lesgourgues~\citep{Gagnon:2011id}, to future work. Although we do
not derive $\delta f_{\mu\nu}$, its likely impact can be bounded
qualitatively. The sub-horizon growth, and with it $D(a)$, $f\sigma_8(z)$
and the linear $P(k)$ shape, together with every background distance that
carries most of the constraining power, follows from the modified expansion
of Eq.~\eqref{eq:background} alone and is insensitive to $\delta f_{\mu\nu}$; the
perturbed entropic sector can influence only the near- and super-horizon
response, namely the late-time ISW contribution at low $\ell$ and the
lensing amplitude. Because this is the same channel governed by the
effective sound speed, the plausible size of the effect is bracketed by the
$c_s^2$ sensitivity analysis of Appendix~\ref{app:cs2}, where scanning the
full clustering-to-smooth range shifts every observable by less than
$1.6\%$, far below cosmic variance and current sensitivity. We therefore
expect the $\delta\zeta=0$ approximation to leave the present posteriors
unchanged, with the residual caveat that a non-local $\delta f_{\mu\nu}$
could source a non-adiabatic pressure not captured by a single $c_s^2$,
whose quantification belongs to the same future derivation.

Because the sub-horizon observables are fixed by the background and GREA
carries no propagating scalar degree of freedom, we model the entropic
component at the linear level as an effective dark-energy fluid with
equation of state $w(a)$ and a prescribed sound speed, in the spirit of the
effective-fluid approach of~\citep{Arjona:2018jhh,Cardona:2020ama}. The background follows exactly from $w(a)$. Under the adopted smooth-fluid approximation, the sub-horizon growth is then entirely determined by the background evolution, while the near- and super-horizon response, which feeds the late-time ISW effect and the lensing spectrum, depends on whether the component clusters.
Pending the derivation of $\delta f_{\mu\nu}$, we adopt the smooth limit
$\cs=1$ throughout, the standard and most conservative choice for a
non-clustering dark-energy fluid. The near- and super-horizon response, and
with it the low-$\ell$ ISW and lensing signals, would be bracketed by also
evolving the clustering limit $\cs=0$; we do not run that case here and
leave its quantification to the same future work that derives
$\delta f_{\mu\nu}$. Since $w(a)$ crosses the phantom divide, where a
perfect-fluid description is singular, the perturbations are evolved with
the parametrized-post-Friedmann scheme~\citep{Fang:2008sn}, which keeps them
regular through $w=-1$ while conserving energy and momentum.

Although the single GREA parameter $\alpha$ is defined through a curvature
scale, the ratio of the causal horizon to the spatial curvature
radius, the model as implemented introduces no geometric curvature of the
spatial sections. The Friedmann equation is solved in the spatially flat
form $H=\sqrt{\rho_{\rm tot}}$, with $\Omega_k\equiv0$, and the entropic
contribution enters exclusively as an effective dark-energy fluid whose
background density is
\begin{equation}
  \rho_{\rm GREA}(a)\;\propto\;\frac{\sinh(2\tau)}{a^{2}}\,,
  \label{eq:rho_grea}
\end{equation}
where $\tau$ denotes the (suitably curvature-normalized) conformal time, with
$\tau_0=\sqrt{-k}\,\eta_0$. The prefactor $a^{-2}$ endows the
component with an intrinsic curvature-like scaling, corresponding to
$w=-1/3$, while the factor $\sinh(2\tau)$ renders it dynamically negligible
in the early Universe ($\sinh 2\tau\simeq2\tau\to0$) and drives the
late-time acceleration as the horizon grows, with $w(a)$ approaching $-1$
today and crossing into the phantom regime. Crucially, because the spatial
geometry remains flat, the primordial power spectrum and the inflationary
predictions are those of a flat cosmology and are left unmodified: the
curvature-like scaling is carried by the effective fluid and evolved
through the PPF prescription~\citep{Garcia-Bellido:2024qau}, not by the
background geometry. GREA therefore remains observationally
distinct from genuine spatial curvature $\Omega_k$: the metric is
flat, so the primordial spectrum and the inflationary predictions are
those of a flat cosmology, and the effective $w(a)$ is time dependent,
reaching the curvature-like value $w=-1/3$ only asymptotically while
crossing into the phantom regime at late times, in contrast with the
static $k/a^2$ term of true curvature. The enhanced growth discussed
above follows from this modified late-time expansion acting on the
matter fluctuations of Eq.~\eqref{eq:growth}, with the entropic
component itself treated as smooth ($\cs=1$) in the baseline.

\section{Implementation in {\sc class}}
\label{sec:implementation}

GREA is not a Horndeski theory: it adds a non-equilibrium boundary term
rather than a propagating scalar, so the $\alpha$-function machinery of
{\sc hi\_class} does not apply. We therefore work in standard
\class~3.3.4~\citep{Blas:2011rf} with its imperfect-fluid dark-energy
module, integrating the GREA background ODE \eqref{eq:background} directly
in the background module rather than feeding a tabulated $w(a)$ through the
standard input, and describing the entropic component at the perturbation
level as the effective fluid of Section~\ref{sec:perturbations}, selected
through a dedicated input option with the PPF closure and the sound speed
fixed as described above.

The main implementation subtlety is the normalization noted below
Eq.~\eqref{eq:alpha}. GREA's time variable $\tau=H_0\eta$ uses a fiducial
$H_0$ that is not the physical present-day rate; the latter is obtained by
integrating \eqref{eq:background} to $a=1$, and in general
$E(z=0)\equiv H(z{=}0)/H_0\neq1$. \class\ instead assumes $H(z{=}0)=H_0$ and
propagates that value through the photon and baryon budget, the sound
horizon $r_s$, and all distances. The implementation rescales the
normalization by $E(z=0)$ so that the physical present-day rate is the one
used by \class, leaving the early-time quantities $\omega_b$,
$\omega_{cdm}$, and $r_s$ untouched. If the rescaling is omitted, \class\
closes the density budget at $a=1$ against the fiducial $H_0$ rather than
the true present-day rate $E(z=0)H_0$, and the physical densities
$\omega_b,\omega_{cdm}$ together with the sound horizon $r_s$ are shifted to
satisfy this constraint, even though GREA is dynamically negligible at the
early times that actually set them. The distortion does not show up in the
dimensionless expansion $E(z)$ or in the shape of the distances, which
remain superficially plausible, but it biases precisely the absolute
early-time calibration that anchors GREA to the CMB, namely $r_s$ and the
acoustic scale $\theta_s=r_s/D_A(z_\ast)$. Because such a normalization
error is silent at the level of the expansion shape and surfaces only in
these calibrated quantities, we verify the rescaling with an explicit
numerical test below.

The model is sampled in the dimensionless combination
$\code{sqrt\_k\_eta0}\equiv\skea=\alpha\,\Dh(z{=}0)$ of
Eq.~\eqref{eq:alpha}, which is what \class\ parses when the GREA equation
of state is active. Recovering the conventional $\alpha$ from a sampled
$\skea$ is not an analytic rescaling but requires inverting the background
mapping $\alpha\mapsto\skea$ with a full cosmological evaluation, which we
do in post-processing. Adiabatic initial conditions apply unchanged, GREA
being subdominant at early times.

\begin{figure}[H]
\centering
\includegraphics[width=1\linewidth]{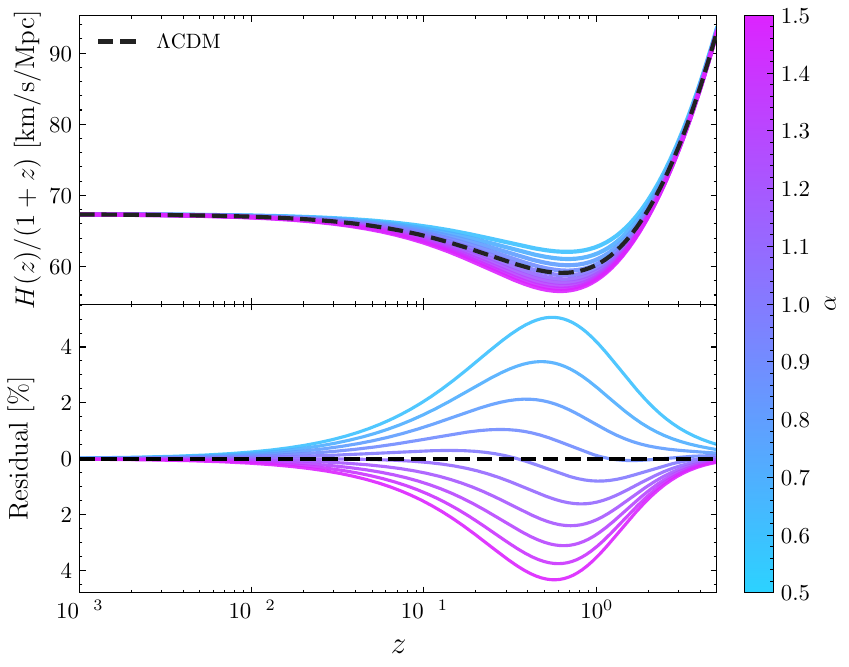}
\caption{Normalized expansion rate $H(z)/(1+z)$ for GREA realizations with varying $\alpha$ (color bar), compared to the \LCDM\ reference (dashed). The lower panel shows the percentage residuals relative to \LCDM.}
\label{fig:H(z)}
\end{figure}

\begin{figure*}[htb!]
\centering
    \begin{minipage}{0.49\textwidth}
        \centering
    \includegraphics[width=1\linewidth]{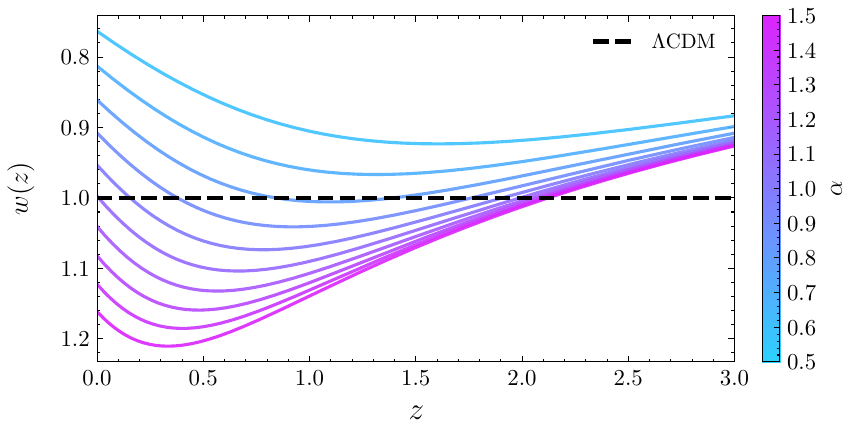}
    \end{minipage}%
    \hfill
    \begin{minipage}[c]{0.49\textwidth}
        \centering
    \includegraphics[width=1\linewidth]{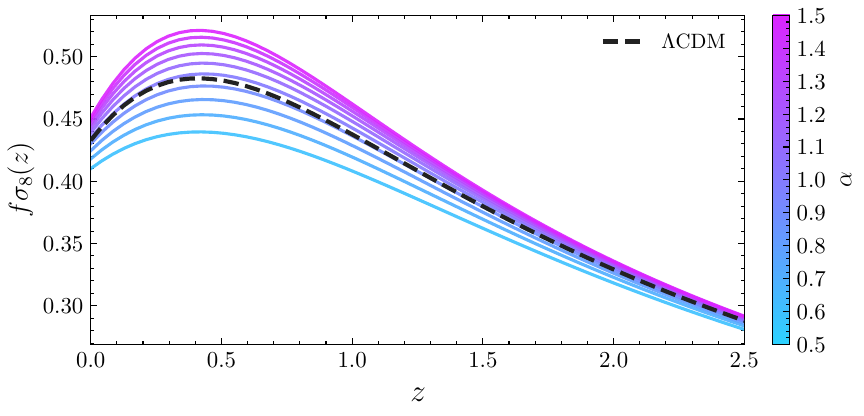}
    \end{minipage}%
    \par
\centering
    \begin{minipage}[t]{0.49\textwidth}
        \centering
        \caption{Effective equation of state $w(z)$ of the entropic component for
        GREA realizations with varying $\alpha$, exhibiting a transient
        phantom crossing at $z\lesssim2$. We point out that most of these configurations have two crossings of the phantom divide.}
        \label{fig:w(z)}
    \end{minipage}%
    \hfill
    \begin{minipage}[t]{0.49\textwidth}
        \centering
        \caption{Linear growth rate $\fsig(z)$ for GREA realizations with varying
        $\alpha$, compared to \LCDM. The reproduction of the semi-analytic
        result of Ref.~\citep{Garcia-Bellido:2024qau} is a validation test;
        the enhanced amplitude at fixed $A_s$ for $\alpha\gtrsim1$
        follows the $S_8$ discussion in Section~\ref{sec:perturbations}.}
\label{fig:growth}
    \end{minipage}
\end{figure*}

\subsection{Validation and Boltzmann-level outputs}
\label{sec:outputs}

We validate the implementation with a harness of quantitative acceptance
tests that must pass before any perturbation output is used, following the
principle that a wrong background should fail first; the tests, together
with the modified code and the run configurations, are available in the
public repository\footnote{The modified \class\ code, the validation harness, and the \cobaya\ input files used in this work are publicly available at \url{https://github.com/zmonneee/CLASS_GREA}.}.

\begin{figure*}[htb!]
\centering
    \begin{minipage}{0.49\textwidth}
        \centering
        \includegraphics[width=1\linewidth]{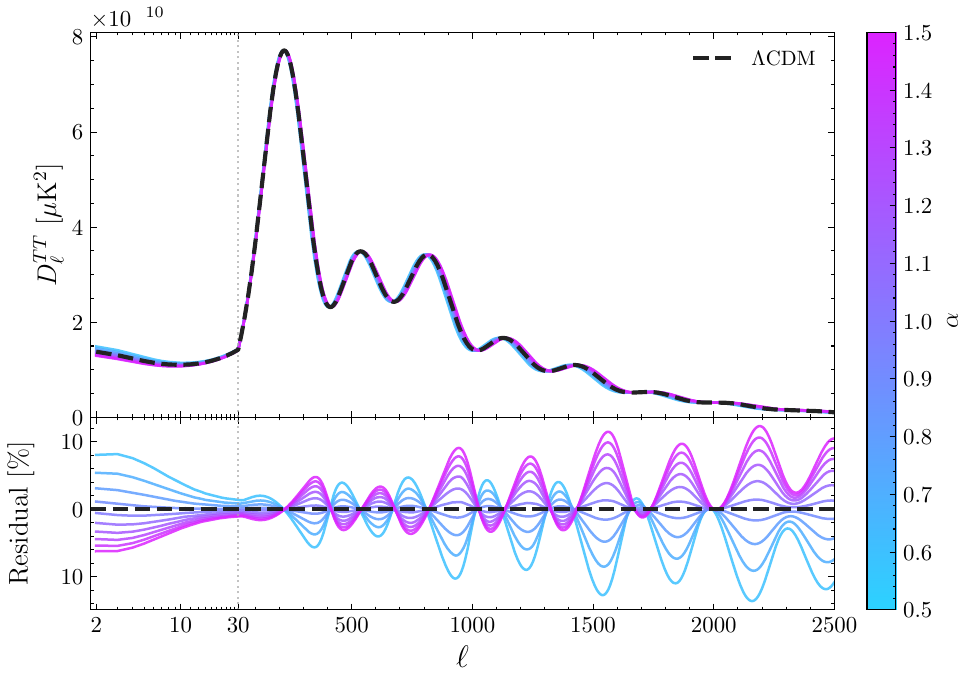}
    \end{minipage}%
    \hfill
    \begin{minipage}[c]{0.49\textwidth}
        \centering
        \includegraphics[width=1\linewidth]{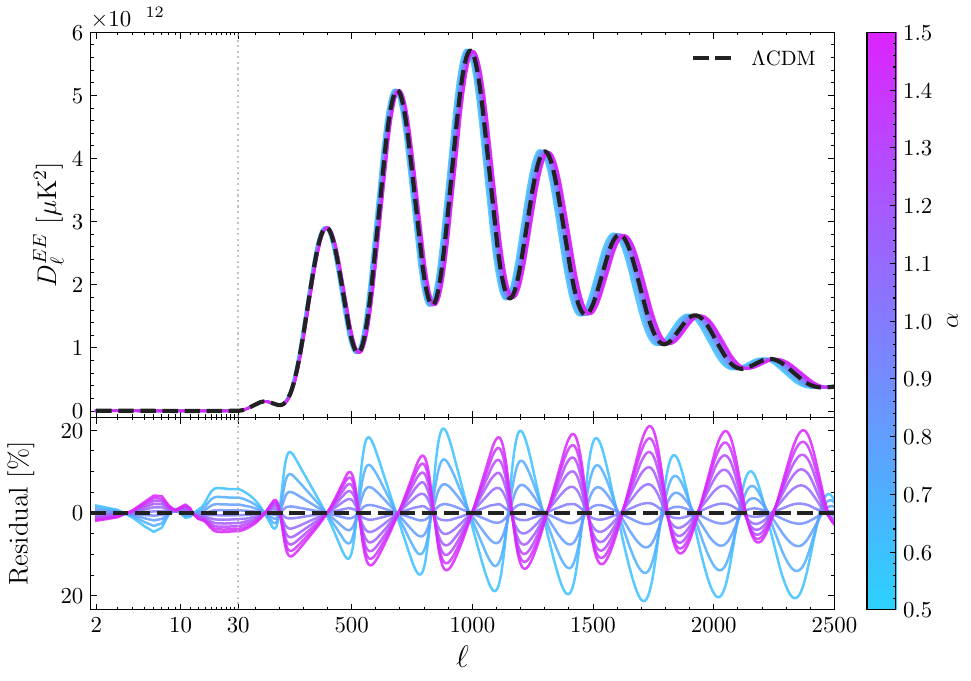}
    \end{minipage}%
    \par
\caption{CMB temperature ({\it Left}) and $E$-mode polarization ({\it Right}) angular
power spectra for GREA realizations with varying $\alpha$, compared to the
\LCDM\ reference (dashed); the lower panels show the percentage
residuals. With the early-time physics held fixed, the modified late-time
expansion shifts the angular scale of the acoustic peaks, producing
oscillatory residuals that grow towards high $\ell$, while the
differences at $\ell\lesssim30$ in $TT$ trace the late-time ISW response
of the entropic fluid.}
\label{fig:power_spectrum_1}
\end{figure*}

\begin{figure*}[htb!]
\centering
    \begin{minipage}{0.49\textwidth}
        \centering
        \includegraphics[width=1\linewidth]{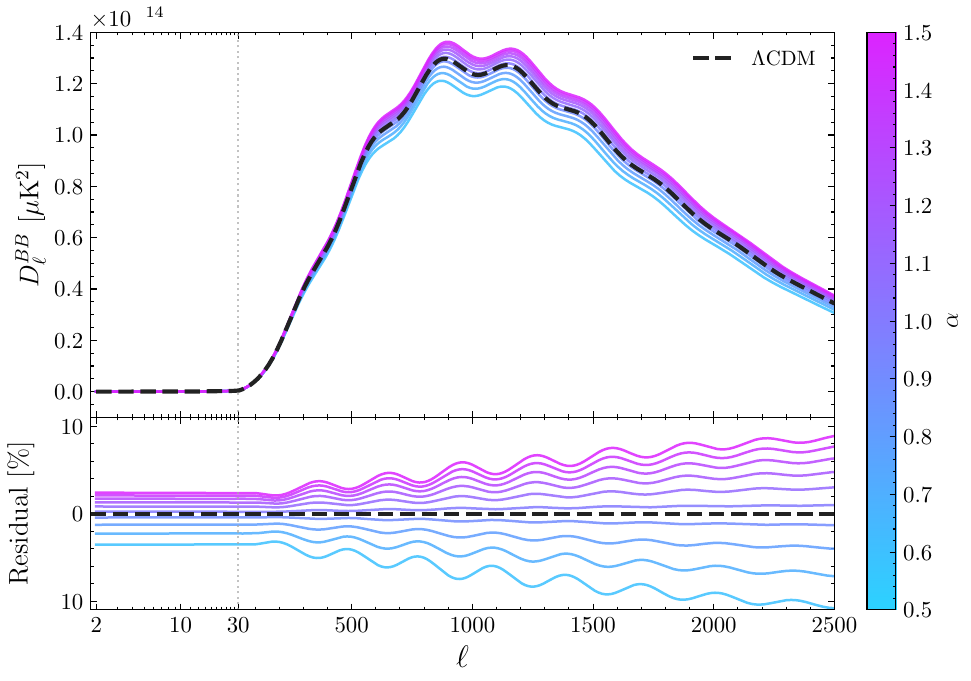}
    \end{minipage}%
    \hfill
    \begin{minipage}[c]{0.49\textwidth}
        \centering
        \includegraphics[width=1\linewidth]{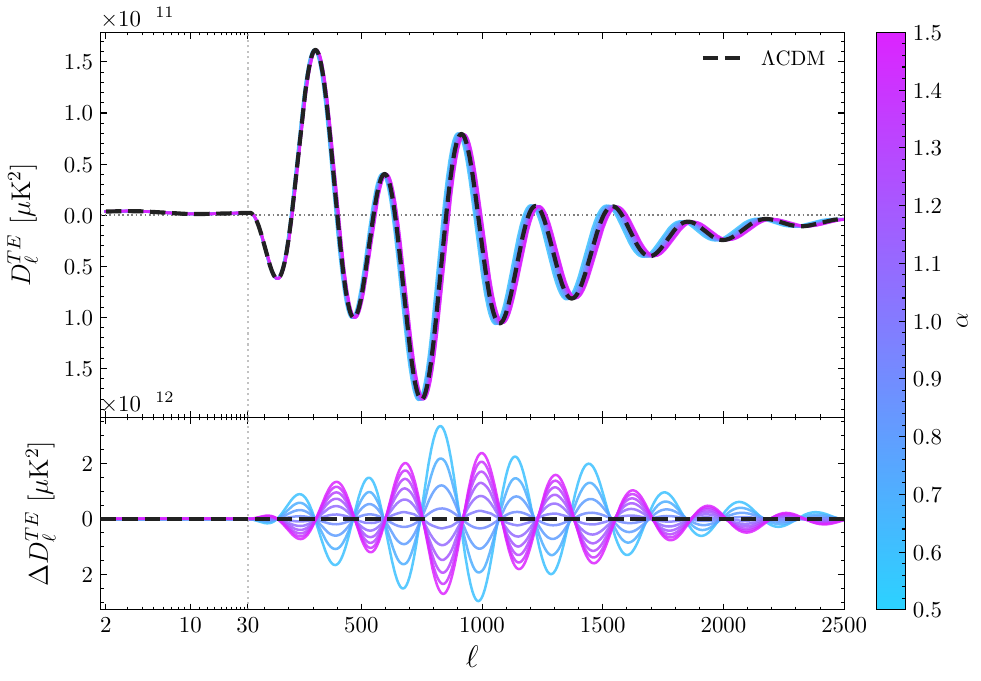}
    \end{minipage}%
    \par
\caption{Same as Fig.~\ref{fig:power_spectrum_1} for the lensing-induced
$B$-mode spectrum ({\it Left}; no primordial tensors) and the $TE$
cross-spectrum ({\it Right}). The $BB$ residuals show a broadband,
nearly scale-independent amplitude shift tracking the modified growth of
structure (cf.\ Fig.~\ref{fig:growth}); the $TE$ residuals are shown as
absolute differences $\Delta D_\ell^{TE}$ because the spectrum crosses
zero.}
\label{fig:power_spectrum_2}
\end{figure*}

Figures~\ref{fig:H(z)}--\ref{fig:power_spectrum_2} collect the validated
outputs for GREA realizations with $\alpha\in[0.5,1.5]$, with the
standard parameters held at the Planck 2018 TT,TE,EE+lowE+lensing \LCDM\
best fit~\citep{Planck:2018vyg},
$(\omega_b,\omega_{cdm},h,A_s,n_s,\tau_{\rm reio})
=(0.02237,\,0.1200,\,0.6736,\,2.1\times10^{-9},\,0.9649,\,0.0544)$,
and only $\alpha$ varied; \LCDM\ at the same parameters is the reference.
Since $h$ is held fixed and the normalization of
Section~\ref{sec:implementation} makes it the physical present-day rate,
all realizations share the same $H(z{=}0)$ and the same early-time
expansion, so the departure from \LCDM\ is confined to intermediate
redshifts. This is visible at the background level
(Fig.~\ref{fig:H(z)}): the residuals vanish at both ends, peak at the
few-percent level around $z\simeq0.5$--$1$, and are ordered monotonically
with $\alpha$, changing sign across $\alpha\simeq1$. The corresponding
effective equation of state (Fig.~\ref{fig:w(z)}) exhibits the transient
phantom phase of Eq.~\eqref{eq:weff}, deepest for the largest $\alpha$.
The growth output (Fig.~\ref{fig:growth}) reproduces the semi-analytic
$\fsig(z)$ of Ref.~\citep{Garcia-Bellido:2024qau}, completing the
validation, with the enhancement at fixed $A_s$ discussed in
Section~\ref{sec:perturbations} directly visible for $\alpha\gtrsim1$.

The angular power spectra are the qualitatively new outputs. A subtlety
in interpreting them is that GREA does not admit a \LCDM\ limit: setting
$\alpha\to0$ yields a matter-dominated, $\Lambda=0$ universe rather than
flat \LCDM, so the implementation cannot be validated by tuning a single
parameter until \class\ reproduces its own standard output. The reference
\LCDM\ curves shown here and throughout are therefore fixed-parameter
comparators evaluated at the same $(\omega_b,\omega_{cdm},\dots)$, not a
limiting case recovered from within GREA, and the validation instead rests
on the parameter-free targets of Sec.~\ref{sec:perturbations}, chiefly the
reproduction of the semi-analytic $f\sigma_8(z)$ of
Ref.~\citep{Garcia-Bellido:2024qau}. With $\omega_b$ and $\omega_{cdm}$
fixed, the sound horizon $r_s$ is common to all realizations, and GREA
affects the primary CMB only through the modified late-time expansion, which
shifts the angular diameter distance to last scattering and hence the
acoustic scale: the result is the oscillatory residual pattern of
Fig.~\ref{fig:power_spectrum_1}, growing towards high $\ell$ and reaching
$\sim10\%$ in $TT$ and $\sim20\%$ in $EE$ at $\ell\simeq2500$ across the
$\alpha$ range shown. The residuals at $\ell\lesssim30$ in $TT$ are of a
different nature, tracing the late-time ISW response of the effective fluid,
and are the part of the signal sensitive to the sound-speed bracket of
Section~\ref{sec:perturbations}. The lensing-induced $B$-mode spectrum
(Fig.~\ref{fig:power_spectrum_2}, with no primordial tensors) shows instead
a broadband, nearly $\ell$-independent amplitude shift, monotonic in
$\alpha$ and tracking the modified growth of Fig.~\ref{fig:growth}; the $TE$
residuals are displayed as absolute differences since the cross-spectrum
changes sign. We stress that these curves illustrate the raw sensitivity to
$\alpha$ at fixed input parameters: in the fits of Section~\ref{sec:data}
the acoustic-scale shift is largely reabsorbed by the recalibration of
$H(z{=}0)$ and $\Omega_{\rm m}$, so the residuals of the best-fit GREA
models are far smaller, and the constraining power comes from the interplay
of the geometric shift with the late-time data and the ISW/lensing signal.

\section{Data and methodology}
\label{sec:data}

We constrain GREA against the benchmark data combinations defined by the
CosmoVerse Cosmology Compilation Group within the CosmoVerse
program~\citep{CosmoVerseNetwork:2025alb}, allowing for a direct comparison
with the standard cosmological analyses performed on the same benchmark.
The benchmark combines the latest observations of the cosmic microwave
background (CMB), baryon acoustic oscillations (BAO), and Type Ia
supernovae (SNIa).

The CMB information is provided by the \textbf{CMB-SPA} likelihood,
which combines measurements from the three major CMB experiments:
\textit{Planck}, ACT and SPT. Specifically, it includes the Planck 2018
low-$\ell$ temperature likelihood (\texttt{Commander}) over the multipole
range $2\le\ell<30$, the Planck 2018 high-$\ell$ TT, TE and EE
likelihood (\texttt{Plik}) using the TT, TE and EE spectra with cuts at
$\ell_{\rm max}=(1000,600,600)$, respectively, the ACT DR6
foreground-marginalized TT, TE and EE likelihood over the multipole range
$\ell\ge600$, and the SPT-3G D1 TT, TE and EE likelihood over the range
$\ell\ge400$, with TT extending to $\ell_{\rm max}=3000$ and TE/EE to
$\ell_{\rm max}=4000$
\citep{Planck:2019nip,ACT:2025fju,SPT-3G:2025bzu}. These are complemented
by the Planck lensing likelihood~\citep{Carron:2022eyg}, the ACT DR6
lensing likelihood~\citep{ACT:2023dou,ACT:2023kun}, the SPT-3G Year-2
\texttt{MUSE} lensing likelihood~\citep{SPT-3G:2024atg}, and a Gaussian
prior on the optical depth to reionization,
$\tau_{\rm reio}=0.051\pm0.006$. We refer to this combined likelihood as
\textbf{CMB-SPA}.

The BAO information is provided by the DESI Data Release 2 measurements,
hereafter referred to as \textbf{DESI}
\citep{DESI:2025zgx}. This data set provides correlated measurements of
the BAO distance scale over the redshift range $0.1<z<2.1$, using more
than 14 million galaxies, quasars and Lyman-$\alpha$ forest tracers,
yielding sub-percent precision constraints on the late-time expansion
history.

For Type Ia supernovae, we consider three complementary samples. The
first is the \textit{Pantheon+} compilation, denoted as \textbf{PP},
consisting of 1701 light curves from 1550 spectroscopically confirmed
Type Ia supernovae spanning the redshift range $0<z\lesssim2.3$
\citep{Brout:2022vxf}. We also consider the Cepheid-calibrated
Pantheon+ sample based on the SH0ES distance ladder~\citep{Riess:2021jrx}, denoted as
\textbf{PPS}, which provides an absolute
calibration of the supernova distances and therefore constrains the local
value of the Hubble constant. Finally, we use the DES Year-5 supernova
sample reanalyzed using the Dovekie photometric calibration,
denoted as \textbf{DD}
\citep{DES:2024jxu,Popovic:2025glk,DES:2025sig}, which provides an
independent late-time distance probe with improved photometric
calibration.

The cosmological constraints presented in this work are obtained from
various combinations of these benchmark data sets. The late-time
combinations constrain the expansion history through BAO and SNIa
distances alone, while the inclusion of CMB-SPA additionally probes the
late-time integrated Sachs--Wolfe effect and CMB lensing, providing the
principal motivation for implementing GREA within an
Einstein--Boltzmann solver.

Posteriors are sampled with \cobaya~\citep{Torrado:2020dgo}, using the
modified \class\ as the theory code. The inputs follow the CCG \class\
templates, with the entropic fluid replacing the cosmological constant and
$\skea$ sampled with a flat prior over the range $[2.5,4.5]$, alongside the
standard cosmological parameters; the full set of priors is given in
Table~\ref{tab:priors}.\footnote{Since GREA rescales the normalization by
$E(z=0)$ so that $H(z{=}0)\neq H_0$, the sampled $H_0$ sets the \class\
normalization $\tau=H_0\eta$ rather than the physical present-day rate; the
latter follows from the $E(z=0)$ rescaling of Sec.~\ref{sec:implementation}
and is the derived $H_0$ reported in Table~\ref{tab:grea_best_fit}.} The full run
configurations are available in the public repository. Each analysis is
performed using four MPI chains with the Metropolis--Hastings sampler, and
convergence is assessed using the Gelman--Rubin statistic, requiring
$R-1<0.02$. The resulting chains are analyzed with
\code{getdist}~\citep{Lewis:2019xzd}, with $\skea$ mapped to $\alpha$ and
the derived parameters $(\Omega_{\rm m},\sigma_8,S_8)$ through the
background inversion described in Section~\ref{sec:implementation}.

\setlength{\arrayrulewidth}{0.1mm}
\setlength{\tabcolsep}{2pt}
\renewcommand{\arraystretch}{1.3}
\begin{table}[t]
\centering
\caption{Priors on the sampled cosmological parameters. All priors are
flat ($\mathcal{U}$) over the quoted range except $\tau_{\rm reio}$, for
which a Gaussian prior $\mathcal{N}(\mu,\sigma)$ is imposed consistently
with the CMB-SPA recipe. The entropic parameter $\skea$ replaces the
cosmological constant; the remaining parameters follow the standard CCG
benchmark. The likelihood nuisance and calibration parameters
($A_{\rm planck}$, $p_{\rm ACT}$, $T_{\rm cal}$, $E_{\rm cal}$) are
sampled with their default likelihood priors and are not listed here.}
\label{tab:priors}
\begin{ruledtabular}
\begin{tabular}{lc}
Parameter & Prior \\
\colrule
$\skea$              & $\mathcal{U}[2.5,\,4.5]$ \\
$\omega_b$           & $\mathcal{U}[0.005,\,0.1]$ \\
$\omega_{cdm}$       & $\mathcal{U}[0.001,\,0.99]$ \\
$H_0$                & $\mathcal{U}[40,\,100]$ \\
$\log(10^{10}A_s)$   & $\mathcal{U}[1.61,\,3.91]$ \\
$n_s$                & $\mathcal{U}[0.8,\,1.2]$ \\
$\tau_{\rm reio}$    & $\mathcal{N}(0.051,\,0.006)$ \\
\end{tabular}
\end{ruledtabular}
\end{table}

For each dataset combination, we assess the performance of GREA relative to \LCDM\ using both the minimum $\chi^2$ and the Bayesian evidence. The former provides a measure of the goodness of fit, while the latter quantifies the relative probability of the competing models after marginalizing over their parameter spaces. Assuming equal prior probabilities for the two models, we define the log-Bayes factor as
\begin{equation}
    \ln\mathcal{B}=\ln\mathcal{Z}_{\rm GREA}-\ln\mathcal{Z}_{\Lambda\rm CDM},
\end{equation}
where
\begin{equation}
    \mathcal{Z}=P(d|\mathcal{M})=\int \mathrm{d}\theta\,\mathcal{L}(\theta)\pi(\theta)
\end{equation}
is the Bayesian evidence, with $\mathcal{L}$ the likelihood and $\pi(\theta)$ the prior distribution. Positive (negative) values of $\ln\mathcal{B}$ indicate a preference for GREA (\LCDM). The Bayesian evidence is computed using the learned harmonic mean estimator (LHME) with normalizing flows, as implemented in the publicly available \texttt{harmonic}\footnote{\url{https://github.com/astro-informatics/harmonic}} package~\citep{mcewenMachineLearningAssisted2023,Polanska:2024arc}. For the analyses presented in this work, we used the \texttt{Python} package \texttt{cosmctools},\footnote{\url{https://github.com/dlehdgk/cosmctools}} which provides a wrapper around \texttt{harmonic} for Bayesian evidence calculations from \texttt{Cobaya} chains, together with a suite of statistical tools for cosmological model comparison.

\section{Results}
\label{sec:results}

\begin{table*}[ht]
\centering
\caption{Marginalized 68\,\% constraints on the GREA cosmological
parameters from the different data combinations. We define
$\Delta\chi^2=\chi^2_{\mathrm{GREA}}-\chi^2_{\Lambda\mathrm{CDM}}$, such
that $\Delta\chi^2>0$ favors $\Lambda$CDM. Similarly, we define
$\ln\mathcal{B}=\ln\mathcal{Z}_{\mathrm{GREA}}-\ln\mathcal{Z}_{\Lambda\mathrm{CDM}}$,
such that $\ln\mathcal{B}<0$ indicates greater support for $\Lambda$CDM.}
\label{tab:grea_best_fit}
\resizebox{\textwidth}{!}{%
\begin{tabular}{lccccccccc}
\hline\hline
Parameter & PP+DESI & PPS+DESI & CMB-SPA & CMB-SPA & CMB-SPA & CMB-SPA & CMB-SPA & CMB-SPA & CMB-SPA \\
 &  &  &  & +DESI & +PP & +PP+DESI & +PPS+DESI & +DD & +DD+DESI \\
\hline
\multicolumn{10}{l}{\textit{Sampled parameters}} \\
$\sqrt{-k}\,\eta_0$ & $3.12^{+0.13}_{-0.14}$ & $3.12^{+0.14}_{-0.13}$ & $3.88^{+0.48}_{-0.50}$ & $3.45 \pm 0.15$ & $3.26 \pm 0.12$ & $3.287^{+0.091}_{-0.092}$ & $3.487^{+0.084}_{-0.083}$ & $3.244^{+0.098}_{-0.101}$ & $3.263 \pm 0.088$ \\
$\log(10^{10} A_\mathrm{s})$ & unconstr. & unconstr. & $3.047^{+0.010}_{-0.011}$ & $3.0569 \pm 0.0099$ & $3.050 \pm 0.010$ & $3.0589^{+0.0097}_{-0.0096}$ & $3.0582^{+0.0096}_{-0.0099}$ & $3.050^{+0.010}_{-0.011}$ & $3.0598^{+0.0097}_{-0.0099}$ \\
$n_\mathrm{s}$ & unconstr. & unconstr. & $0.9694^{+0.0033}_{-0.0032}$ & $0.9725 \pm 0.0030$ & $0.9688 \pm 0.0032$ & $0.9732 \pm 0.0029$ & $0.9730^{+0.0030}_{-0.0029}$ & $0.9688^{+0.0033}_{-0.0032}$ & $0.9732 \pm 0.0029$ \\
$\Omega_\mathrm{b} h^2$ & $0.026^{+0.010}_{-0.011}$ & $0.0304 \pm 0.0020$ & $0.022406 \pm 0.000096$ & $0.022453^{+0.000092}_{-0.000091}$ & $0.022396^{+0.000093}_{-0.000094}$ & $0.022460^{+0.000091}_{-0.000092}$ & $0.022486^{+0.000090}_{-0.000093}$ & $0.022393^{+0.000096}_{-0.000094}$ & $0.022465 \pm 0.000093$ \\
$\Omega_\mathrm{c} h^2$ & $0.125^{+0.025}_{-0.026}$ & $0.1339 \pm 0.0059$ & $0.12022^{+0.00099}_{-0.00103}$ & $0.11880 \pm 0.00071$ & $0.12042^{+0.00093}_{-0.00091}$ & $0.11845^{+0.00066}_{-0.00065}$ & $0.11852^{+0.00064}_{-0.00065}$ & $0.12045^{+0.00097}_{-0.00096}$ & $0.11838^{+0.00068}_{-0.00067}$ \\
$H_0 \, [\mathrm{km\,s^{-1}\,Mpc^{-1}}]$ & $70.0^{+8.0}_{-8.7}$ & $73.4 \pm 1.0$ & $71.4^{+3.3}_{-3.6}$ & $68.80^{+0.93}_{-0.95}$ & $66.99 \pm 0.77$ & $67.83 \pm 0.56$ & $69.19^{+0.53}_{-0.52}$ & $66.89^{+0.67}_{-0.68}$ & $67.69^{+0.53}_{-0.54}$ \\
$\tau_\mathrm{reio}$ & unconstr. & unconstr. & $0.0542 \pm 0.0056$ & $0.0580^{+0.0054}_{-0.0053}$ & $0.0547^{+0.0056}_{-0.0057}$ & $0.0590 \pm 0.0052$ & $0.0586^{+0.0052}_{-0.0053}$ & $0.0546^{+0.0056}_{-0.0057}$ & $0.0591 \pm 0.0055$ \\
\hline
\multicolumn{10}{l}{\textit{Derived parameters}} \\
$\alpha$ & $0.967^{+0.041}_{-0.042}$ & $0.965 \pm 0.042$ & $1.20^{+0.15}_{-0.16}$ & $1.067 \pm 0.047$ & $1.008 \pm 0.036$ & $1.018 \pm 0.028$ & $1.079 \pm 0.026$ & $1.004^{+0.030}_{-0.031}$ & $1.010 \pm 0.027$ \\
$\Omega_\mathrm{m}$ & $0.3064 \pm 0.0077$ & $0.3065^{+0.0079}_{-0.0078}$ & $0.283^{+0.029}_{-0.027}$ & $0.2999^{+0.0079}_{-0.0078}$ & $0.3198 \pm 0.0080$ & $0.3078^{+0.0052}_{-0.0051}$ & $0.2959 \pm 0.0046$ & $0.3208^{+0.0073}_{-0.0074}$ & $0.3089^{+0.0049}_{-0.0050}$ \\
$\sigma_8$ & unconstr. & unconstr. & $0.849^{+0.029}_{-0.031}$ & $0.821 \pm 0.011$ & $0.8119^{+0.0081}_{-0.0079}$ & $0.8106 \pm 0.0072$ & $0.8230 \pm 0.0067$ & $0.8111^{+0.0070}_{-0.0072}$ & $0.8091^{+0.0069}_{-0.0068}$ \\
$S_8$ & unconstr. & unconstr. & $0.823^{+0.015}_{-0.014}$ & $0.8208^{+0.0062}_{-0.0064}$ & $0.8381^{+0.0085}_{-0.0084}$ & $0.8210^{+0.0062}_{-0.0060}$ & $0.8173^{+0.0062}_{-0.0061}$ & $0.8387^{+0.0089}_{-0.0087}$ & $0.8210^{+0.0065}_{-0.0064}$ \\
$r_\mathrm{drag}$ & $144^{+18}_{-16}$ & $135.8 \pm 2.1$ & $146.93 \pm 0.25$ & $147.26^{+0.20}_{-0.19}$ & $146.89 \pm 0.24$ & $147.34 \pm 0.19$ & $147.30^{+0.19}_{-0.18}$ & $146.89^{+0.25}_{-0.24}$ & $147.35 \pm 0.19$ \\
\hline
$\chi^2_{\rm min}$ & $1416.1$ & $1465.3$ & $586.6$ & $605.3$ & $1991.2$ & $2009.3$ & $2085.3$ & $2218.6$ & $2237.1$ \\
$\Delta\chi^2$ & $-4.724$ & $-4.707$ & $-0.995$ & $-1.081$ & $-0.048$ & $-2.286$ & $-3.413$ & $-1.404$ & $-6.198$ \\
$\ln\mathcal{B}$ & $0.47\pm0.02$ & $0.58\pm0.03$ & $1.01_{-0.51}^{+0.36}$ & $-2.72^{+1.26}_{-0.55}$ & $-2.33^{+1.04}_{-0.55}$ & $-0.31_{-0.39}^{+0.42}$ & $1.35^{+0.26}_{-0.21}$ & $-0.58\pm0.26$ & $-1.41_{-0.69}^{+2.59}$\\
\hline\hline
\end{tabular}%
}
\end{table*}

\begin{figure}[ht]
\centering
\includegraphics[width=\columnwidth]{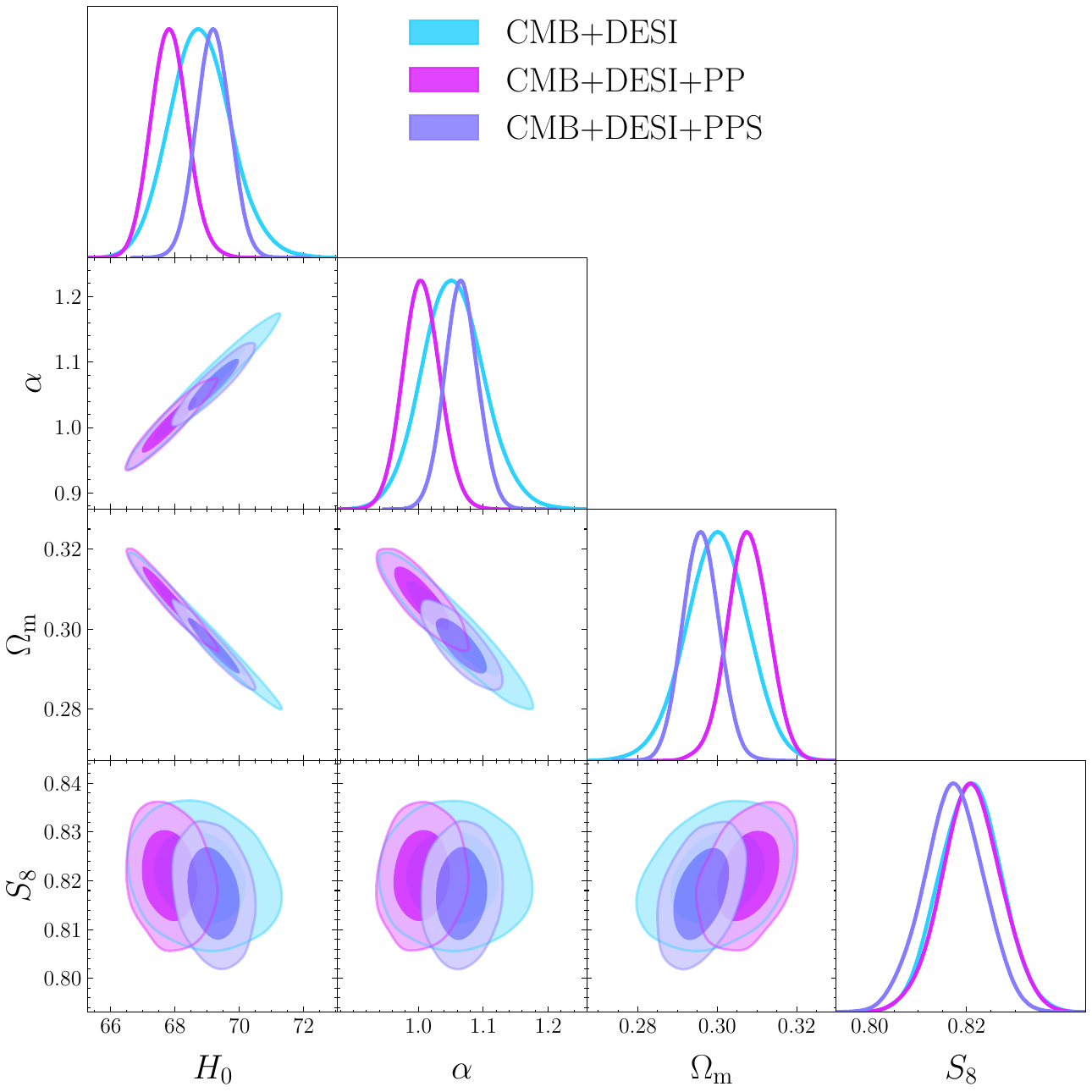}
\caption{Marginalized posteriors for the headline parameters $\{H_0,\,\Omega_m,\, S_8,\,\alpha\}$ across the three CMB-anchored dataset combinations. The GREA coupling $\alpha$ acquires a mild preference for $\alpha > 1$ once geometric late-time probes are added; $S_8$ remains insensitive to the calibration choice.}
\label{fig:headline_triangle}
\end{figure}

\begin{figure}[ht]
\centering
\includegraphics[width=\columnwidth]{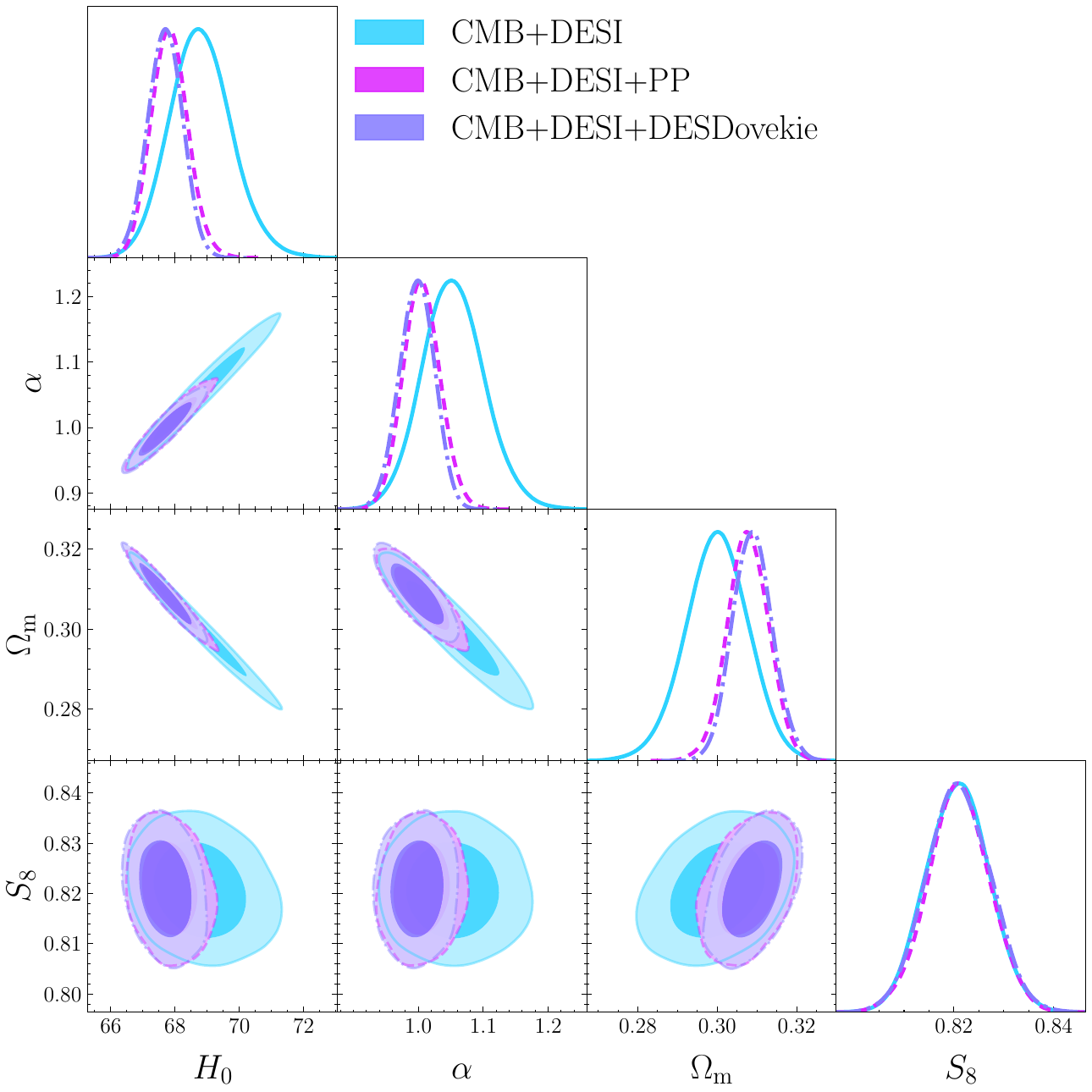}
\caption{Marginalized posteriors for the headline parameters $\{H_0,\,\Omega_m,\, S_8,\,\alpha\}$. Comparison between different Supernovae catalogs, Pantheon+ and DES Dovekie.}
\label{fig:headline_triangle_Sne}
\end{figure}

\begin{figure}[ht]
\centering
\includegraphics[width=\columnwidth]{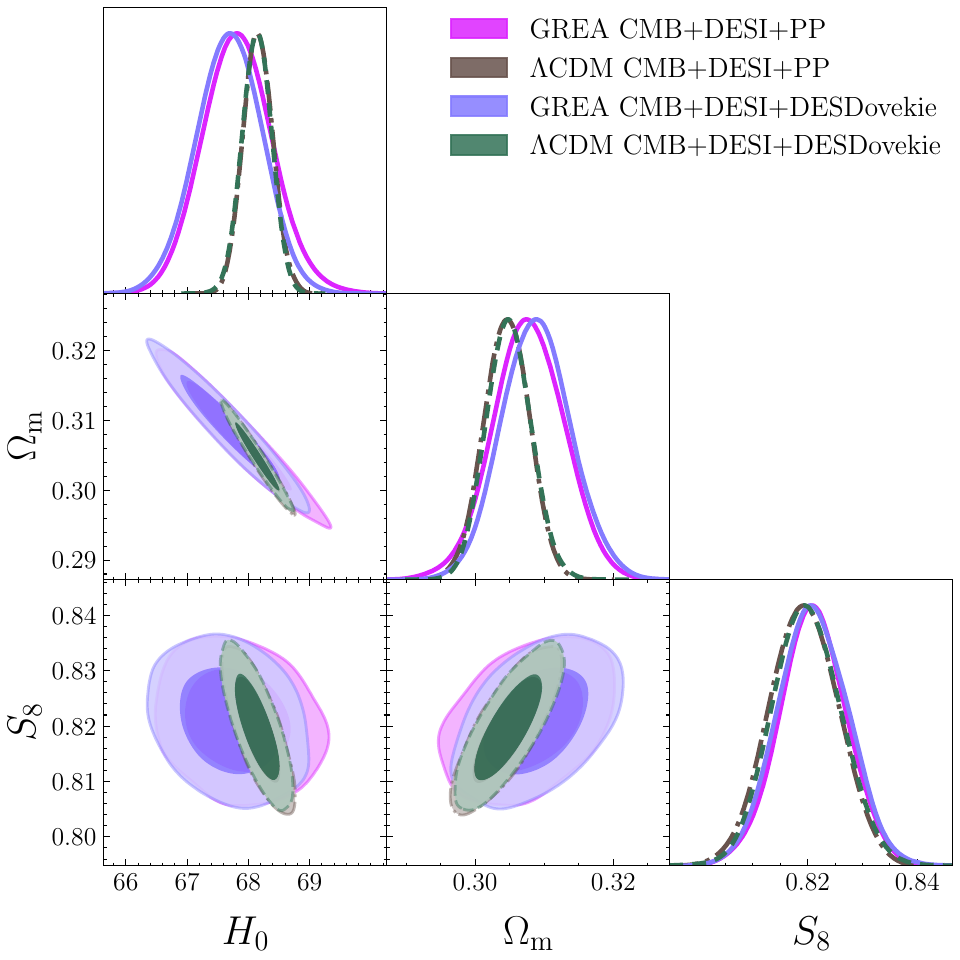}
\caption{Marginalized posteriors for the headline parameters $\{H_0,\,\Omega_m,\, S_8\}$ across the two CMB-anchored dataset combinations. Comparison between different Supernovae catalogs, Pantheon+ and DES Dovekie, and the $\Lambda$CDM model.}
\label{fig:headline_triangle_LCDM}
\end{figure}


We now present the cosmological constraints on the GREA model obtained from the data combinations described in the previous section. The corresponding marginalized constraints are reported in Table~\ref{tab:grea_best_fit}, where we quote both the sampled and derived cosmological parameters, together with the minimum $\chi^2$ and the Bayesian evidence for each data combination. Throughout this work, we compare the performance of GREA with the standard \LCDM\ model using the difference in the best-fit $\chi^2$ and the Bayes factor, whose definitions are given in the table caption.
The corresponding marginalized posterior distributions are shown in Figs.~\ref{fig:headline_triangle} and \ref{fig:headline_triangle_Sne}, illustrating the correlations between the GREA parameter $\alpha$ and the derived cosmological parameters of greatest interest for the different combinations of CMB, BAO and Type Ia supernova data. A direct comparison between GREA and \LCDM\ for the key cosmological parameters $H_0$, $\Omega_m$ and $S_8$ is presented in Fig.~\ref{fig:headline_triangle_LCDM}. The corresponding posterior distributions for the remaining cosmological parameters and data combinations are shown in Appendix \ref{app:additional}. We then present the reconstructed expansion history and effective equation of state in Fig.~\ref{fig:best_Hz_wz}, followed by the comparison with the DESI DR2 BAO distance measurements in Fig.~\ref{fig:best_distances}. 

Focusing first on the background data alone, GREA provides an excellent description of the late-time observations. Both the PP+DESI and PPS+DESI combinations yield a modest improvement in the best-fit $\chi^2$ with respect to \LCDM, while the Bayesian evidence remains statistically inconclusive, indicating that the two models provide an essentially equivalent description of the current background data. However, these data combinations leave strong degeneracies among the cosmological parameters, resulting in broad posterior distributions. Consequently, no robust conclusions can yet be drawn regarding the preferred values of the cosmological parameters or the GREA parameter $\alpha$.

A similar picture emerges when considering the CMB-SPA data alone. GREA provides a fit to the full CMB likelihood that is statistically indistinguishable from \LCDM, with $\Delta\chi^2\simeq-1$ and an inconclusive Bayes factor. At the same time, the preferred cosmological parameters shift in an interesting direction: the inferred Hubble constant increases to $H_0=71.4^{+3.3}_{-3.6}\,\mathrm{km\,s^{-1}\,Mpc^{-1}}$, while the matter density decreases to $\Omega_m=0.283^{+0.029}_{-0.027}$. The GREA parameter is constrained to $\alpha=1.20^{+0.15}_{-0.16}$, fully consistent with the theoretical expectation of $\alpha\sim1$, while the predicted value of $S_8$ remains essentially unchanged with respect to \LCDM. Although these trends are encouraging, they are driven by broad parameter degeneracies within the CMB data alone. Breaking these degeneracies therefore requires the inclusion of complementary late-time distance measurements, to which we now turn.

The addition of the DESI BAO measurements to CMB-SPA substantially reduces the parameter degeneracies present in the CMB-only analysis. In particular, BAO provide a direct measurement of the late-time matter density, favouring a larger value of $\Omega_m$ than preferred by the CMB alone. Since the CMB primarily constrains the physical matter density, $\Omega_m h^2$, the increase in $\Omega_m$ naturally translates into a lower value of $H_0$, reducing the ability of GREA to alleviate the Hubble tension when BAO data are included. The increase in $\Omega_m$ is also accompanied by a decrease in the preferred value of the GREA parameter, reflecting the anticorrelation between $\Omega_m$ and $\alpha$. As a result, the posterior shifts towards the theoretically expected value, with $\alpha=1.067\pm0.047$. The remaining cosmological parameters are largely unaffected by the inclusion of the BAO data.

Adding Type Ia supernovae to the CMB further strengthens this picture. Both the Pantheon+ and DES Dovekie compilations favour a slightly higher matter density than the CMB alone, leading to a further reduction of the inferred Hubble constant. At the same time, the preferred value of the GREA parameter moves even closer to the theoretical prediction, yielding $\alpha=1.008\pm0.036$ for CMB-SPA+PP and $\alpha=1.004^{+0.030}_{-0.031}$ for CMB-SPA+DD, fully consistent with $\alpha=1$ at the $1\sigma$ level. The corresponding increase in $\Omega_m$ also raises the inferred value of $S_8$, potentially worsening the discrepancy with weak-lensing measurements. However, such a conclusion cannot yet be drawn quantitatively, since the weak-lensing constraints on $S_8$ are themselves model dependent and have so far been derived assuming \LCDM. A robust assessment of the $S_8$ tension therefore requires a dedicated analysis of the weak-lensing data within the GREA framework.

Finally, we consider the full combinations of CMB, BAO and Type Ia supernova data. As expected, the CMB-SPA+DESI+PP constraints represent a compromise between the CMB-SPA+DESI and CMB-SPA+PP results. The matter density and Hubble constant are both tightly constrained, with $\Omega_m$ and $H_0$ taking intermediate values between those preferred by the individual data combinations. The same overall picture is recovered when replacing Pantheon+ with the recalibrated DES Dovekie sample, demonstrating that the cosmological constraints are robust with respect to the choice of supernova compilation.

Remarkably, the preferred value of the GREA parameter remains extremely stable across all combined analyses, consistently yielding $\alpha\simeq1$ within the statistical uncertainties, in excellent agreement with the theoretical prediction of the model. At the same time, GREA continues to provide a slightly better best-fit than \LCDM, while the Bayesian evidence remains statistically indistinguishable between the two models.

A comparison with the corresponding \LCDM\ constraints, shown in Fig.~\ref{fig:headline_triangle_LCDM}, reveals that both models recover nearly identical values for the standard cosmological parameters. The main difference lies in the parameter uncertainties, which are typically enlarged by almost a factor of two in GREA. This broadening originates primarily from the strong correlation between $\alpha$, $\Omega_m$ and $H_0$. In contrast, $\alpha$ exhibits little or no correlation with the remaining cosmological parameters, as illustrated in Fig.~\ref{fig:full_triangle}, explaining why their posterior distributions remain essentially unchanged with respect to \LCDM.

For completeness, we also report the constraints obtained when combining the CMB with the SH0ES-calibrated Pantheon+ sample. However, this combination should not be regarded as a statistically consistent data set, since the CMB and SH0ES measurements remain in excess of $3\sigma$ tension within GREA. As in the case of \LCDM, combining inconsistent data sets mainly serves as an illustration of the parameter shifts induced by the local distance-ladder calibration, rather than providing meaningful cosmological constraints.

We now turn to the reconstructed background evolution shown in Fig.~\ref{fig:best_Hz_wz}. The upper panel displays the normalized expansion rate, $H(z)/(1+z)$, for the GREA best fit compared with the corresponding \LCDM\ prediction. Although the two models provide a very similar overall expansion history, GREA predicts a characteristic localized deviation at intermediate redshifts. In particular, the expansion rate is higher than in \LCDM\ over the approximate interval $0.2\lesssim z\lesssim0.6$, before becoming slightly lower at higher redshifts and gradually converging back to the \LCDM\ prediction by $z\simeq2$. This behavior reflects the modified late-time dynamics of the entropic component while preserving the successful early-time expansion history required by the CMB.

The lower panel shows the corresponding effective equation of state of the entropic component. As already anticipated by the parameter constraints, the present-day Universe is in a quintessence-like phase ($w>-1$), followed by a first crossing of the phantom divide at $z\simeq0.3$. This behavior is remarkably similar to that inferred from phenomenological dark-energy parameterizations constrained by current CMB, BAO and Type Ia supernova observations. Unlike these phenomenological models, however, GREA predicts a second crossing of the phantom divide at $z\simeq2$, after which the equation of state returns to the quintessence regime. This second transition is a distinctive prediction of the model, arising naturally from the underlying thermodynamic dynamics rather than from an imposed parametrization.

\begin{figure*}[htb!]
\centering
\includegraphics[width=0.8\textwidth]{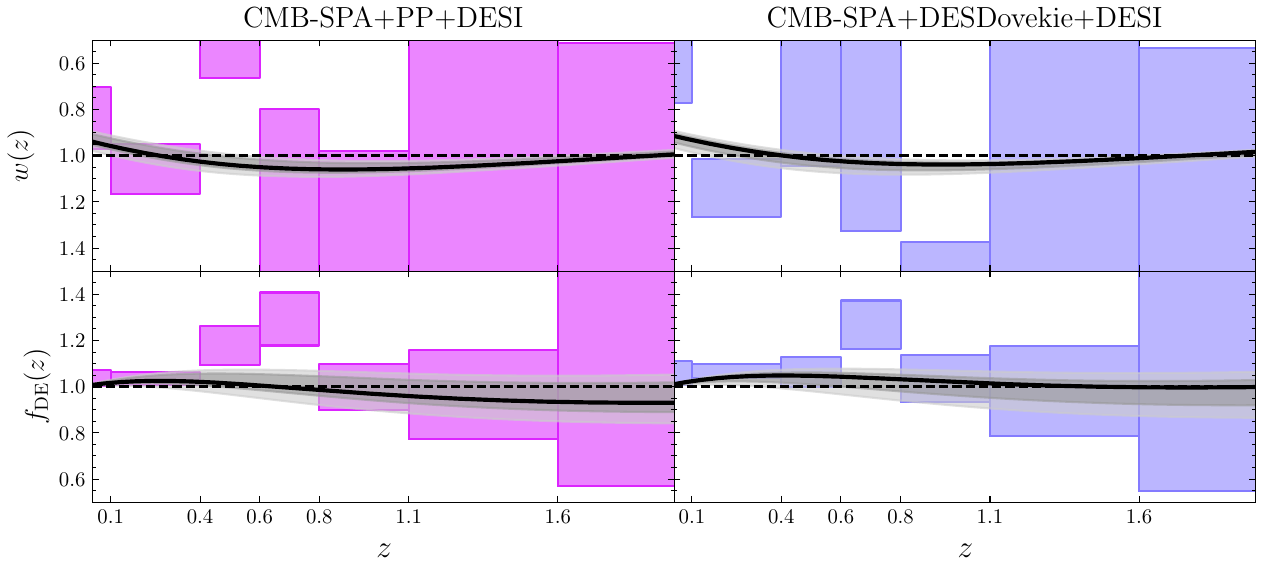}
\caption{\textit{Top}: effective dark-energy equation of state $w(z)$ for the
GREA best fit (solid curves), compared with the binned, model-independent
reconstruction of Ref.~\cite{Kessler:2026dbi} (shaded rectangles). \textit{Bottom}:
the corresponding normalized dark-energy density $f_{\rm DE}(z)$, defined as in Eq.~\eqref{eq:f_def}. Results are shown for the
PP+DESI (magenta) and DD+DESI (purple) combinations. The rectangles give the
$68\%$ credible interval of the reconstruction in each redshift bin, with their
horizontal extent marking the bin width. The GREA curves are evaluated at the best
fit of the same data combination reported in Table~\ref{tab:grea_best_fit}, with
the gray bands showing the $68\%$ and $95\%$ credible intervals. The horizontal dashed lines mark the cosmological-constant values.}
\label{fig:wz_recon}
\end{figure*}

\begin{figure*}[ht]
\centering
    \begin{minipage}{0.49\textwidth}
        \centering
    \includegraphics[width=1\linewidth]{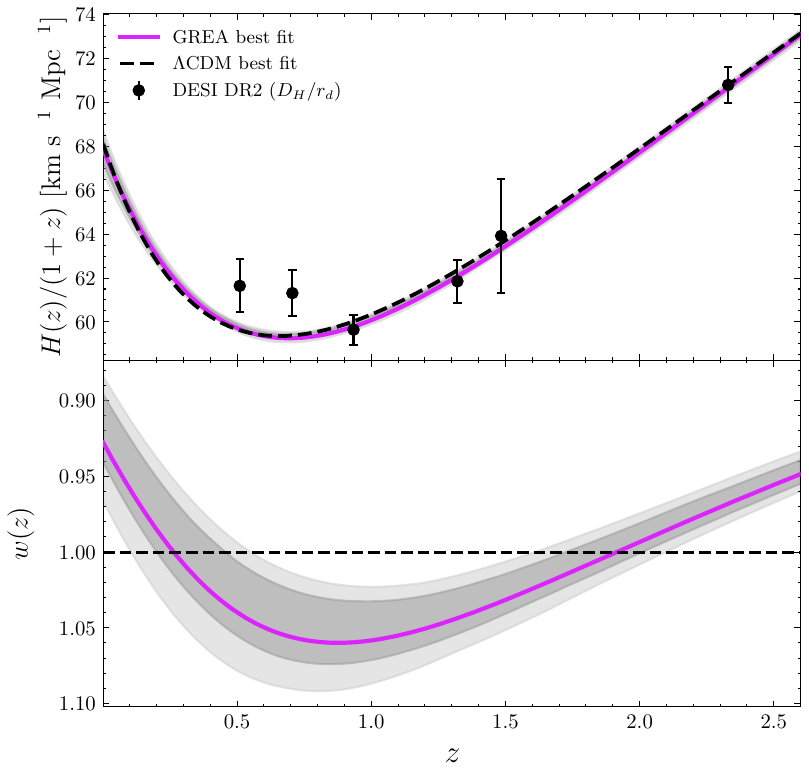}
    \end{minipage}%
    \hfill
    \begin{minipage}[c]{0.49\textwidth}
        \centering
    \includegraphics[width=1\linewidth]{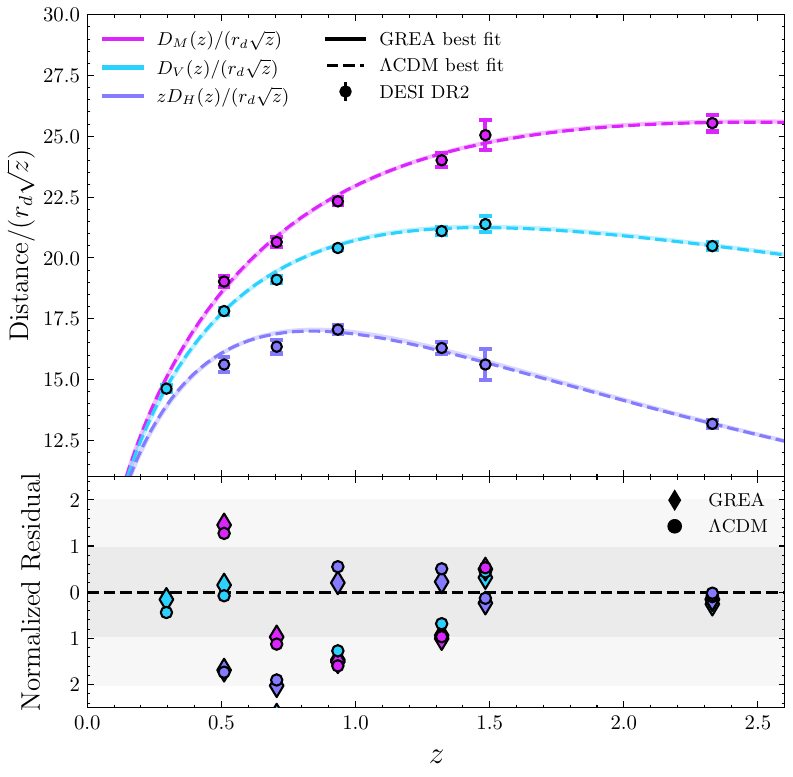}
    \end{minipage}%
    \par
\centering
    \begin{minipage}[t]{0.49\textwidth}
        \centering
        \caption{\textit{Top}: Normalized expansion rate $H(z)/(1+z)$ for the GREA
best fit (magenta) and the \LCDM\ best fit (black dashed), compared with
the DESI DR2 BAO measurements of $D_H/r_d$ (black points with $1\sigma$
error bars). The shaded band indicates the $68\%$ confidence region of
the GREA reconstruction. \textit{Bottom}: Corresponding effective
dark-energy equation of state $w(z)$ for GREA (magenta), with $68\%$ and
$95\%$ confidence regions (dark and light gray). The horizontal dashed
line marks the cosmological-constant value, $w=-1$.}
\label{fig:best_Hz_wz}
    \end{minipage}%
    \hfill
    \begin{minipage}[t]{0.49\textwidth}
        \centering
    \caption{\textit{Top}: DESI DR2 BAO distance summary statistics, scaled by
$r_d\sqrt{z}$: the transverse comoving distance $D_M(z)/(r_d\sqrt{z})$
(magenta), the angle-averaged distance $D_V(z)/(r_d\sqrt{z})$ (cyan), and
$z\,D_H(z)/(r_d\sqrt{z})$ (purple). Solid curves show the GREA best fit,
and dashed curves the \LCDM\ best fit, with the DESI DR2 measurements
overplotted as points with $1\sigma$ error bars. \textit{Bottom}:
Normalized residuals $(\mathrm{data}-\mathrm{model})/\sigma$ for GREA
(diamonds) and \LCDM\ (circles), color-coded by distance measure as in
the upper panel. The dark and light gray bands denote the $1\sigma$ and
$2\sigma$ regions.}
\label{fig:best_distances}
    \end{minipage}
\end{figure*}

A sharper comparison is offered by model-independent reconstructions.
Figure~\ref{fig:wz_recon} sets the GREA prediction against the binned
reconstruction of the dark-energy sector obtained in Ref.~\cite{Kessler:2026dbi}
from the same data combinations. That analysis reconstructs the equation of
state independently in each redshift bin, with no functional form imposed on
$w(z)$, and therefore provides an agnostic target against which a
single-parameter prediction can be tested. GREA follows the reconstructed bins
over most of the redshift range probed by the data, recovering both the
present-day quintessence-like value and the phantom excursion at intermediate
redshift with no freedom beyond $\alpha$. The two panels are not independent,
since $f_{\rm DE}(z)$ is the integral of $w(z)$ through Eq.~\eqref{eq:f_def}, and
the density panel is correspondingly the smoother and more constraining of the
two. The agreement is closest in the lower bins, where the reconstruction is
tightest, while the highest bins widen and retain little discriminating power.
We note that the second phantom crossing predicted by GREA at $z \simeq 2$ lies
beyond the last reconstructed bin and is therefore not probed by the present
data, although it remains the cleanest target for a future extension of this
reconstruction to higher redshift.

Figure~\ref{fig:best_distances} compares the best-fit GREA and \LCDM\ predictions for the DESI DR2 BAO distance measurements. The two models are virtually indistinguishable over the full redshift range probed by the observations, with both accurately reproducing the transverse, radial and volume-averaged BAO distances. This agreement is reflected in the normalized residuals, which remain within the observational uncertainties for both models, explaining the nearly identical goodness of fit found in the statistical analysis.

\subsection{Interpreting the Bayes factor for non-nested models}
\label{app:bayes}

Table~\ref{tab:grea_best_fit} quotes, for each dataset combination, the
$\chi^2$ difference $\Delta\chi^2=\chi^2_{\rm GREA}-\chi^2_{\Lambda\rm CDM}$
and the log-Bayes factor $\ln\mathcal{B}=\ln\mathcal{Z}_{\rm
GREA}-\ln\mathcal{Z}_{\Lambda\rm CDM}$. Both quantities must be interpreted
with care because GREA and \LCDM\ are {\it not} nested: neither model is
recovered as a limiting case of the other. The familiar intuition that a
more flexible model can only improve the fit therefore does not apply, and a
$\Delta\chi^2$ close to zero already carries a strong message, namely that a
single-parameter, thermodynamically predicted expansion history reproduces
the data as well as the finely tuned cosmological
constant~\citep{Calderon:2025dhj}. Across the nine dataset combinations,
the fit differences reported in Table~\ref{tab:grea_best_fit} remain within
$|\Delta\chi^2|\lesssim 6$ despite fitting hundreds to thousands of data
points, indicating that the two models provide statistically
near-indistinguishable fits.

The Bayes factor compares the evidences
$\mathcal{Z}_i=\mathcal{P}(D|M_i)$ and, through Bayes' theorem, is related to
the posterior model odds via the prior model probabilities $\Pi(M_i)$,
\begin{equation}
  \mathcal{B}=\frac{\mathcal{Z}_{\rm GREA}}{\mathcal{Z}_{\Lambda\rm CDM}}
   =\frac{\mathcal{P}(M_{\rm GREA}\,|\,D)}{\mathcal{P}(M_{\Lambda\rm CDM}\,|\,D)}\,
    \frac{\Pi(M_{\Lambda\rm CDM})}{\Pi(M_{\rm GREA})}\,.
  \label{eq:bayes}
\end{equation}
The values reported in Table~\ref{tab:grea_best_fit} assume equal model
priors, $\Pi(M_{\rm GREA})=\Pi(M_{\Lambda\rm CDM})=1/2$, corresponding to
the natural default in which the competing models are regarded as equally
motivated phenomenological descriptions. As stressed in the companion
background analysis~\citep{Calderon:2025dhj}, this assumption is itself open
to debate in the present context: a cosmological constant fine-tuned by
roughly $120$ orders of magnitude with respect to its natural
quantum-field-theory scale arguably deserves a lower prior weight than an
acceleration mechanism derived from horizon thermodynamics. Adopting such
theoretically motivated priors would therefore shift the posterior odds of
Eq.~\eqref{eq:bayes} in favor of GREA. Even under equal priors, however,
the largest is $|\ln\mathcal{B}|\simeq 2.72$ and are
frequently positive, corresponding to evidence ranging from
``inconclusive'' to at most ``moderate support'' on the Jeffreys scale. Moreover, the
usefulness of fixed Jeffreys-scale thresholds is itself known to be limited,
particularly for comparisons between non-nested
models~\citep{Nesseris:2012cq,Keeley:2021dmx,Koo:2021suo}. We therefore
present both $\Delta\chi^2$ and $\ln\mathcal{B}$ as useful summary
statistics rather than definitive model-selection criteria, and emphasize
that the strongest discrimination between GREA and \LCDM\ is expected to
come from growth, ISW and lensing observables, discussed in
Appendix~\ref{sec:grea_vs_genentropy} and
Section~\ref{sec:perturbations}, rather than from the present
background goodness-of-fit statistics.

\section{Discussion and Conclusions}
\label{sec:conclusions}

We have brought General Relativistic Entropic Acceleration from the
background into the realm of cosmological perturbations, presenting the
first implementation of GREA within an Einstein--Boltzmann code. The GREA
background of Eq.~\eqref{eq:background} is integrated directly into
\class, while the entropic component is evolved at the linear level as an
effective dark-energy fluid whose equation of state,
Eq.~\eqref{eq:weff}, is regulated through the parametrized-post-Friedmann
scheme, ensuring that the perturbations remain regular across the transient
phantom crossings. This implementation provides the complete set of
cosmological observables required for a full likelihood analysis, including
the $TT$, $TE$, $EE$, lensing and lensing-induced $BB$ angular power
spectra, the linear matter power spectrum, and the growth rate
$\fsig(z)$, all computed self-consistently rather than through compressed
distance priors. Coupled to \cobaya, the modified code enabled a
Markov-chain Monte Carlo analysis of the full primary-CMB likelihoods from
the CMB-SPA combination together with DESI DR2 BAO and the Pantheon+ and
recalibrated DES Dovekie Type Ia supernova compilations, including
Pantheon+ with the SH0ES calibration. The modified code, validation suite,
and run configurations are publicly released.

The central result is a striking vindication of GREA's defining
prediction. Across every dataset combination, the inferred coupling clusters
tightly around unity, from $\alpha=0.967\pm0.041$ for the late-time
PP+DESI combination to $\alpha=1.079\pm0.026$ for the full
CMB-SPA+PPS+DESI combination (Table~\ref{tab:grea_best_fit},
Fig.~\ref{fig:headline_triangle}). That the size of the causal horizon
today should equal the spatial-curvature scale, $\alpha\sim1$, is not a
fitted outcome but a genuine prediction of the theory, and the data single
it out at the few-percent level. The reconstructed expansion history and
its effective equation of state (Fig.~\ref{fig:best_Hz_wz}) reproduce the
transient phantom crossing anticipated by the model and exhibit a second
crossing around $z\sim2$, while tracing the DESI DR2 BAO distance
measurements as closely as \LCDM\ and dispensing entirely with a
cosmological constant.

This prediction can be confronted directly with agnostic reconstructions of the
dark-energy sector. Compared with the binned reconstruction of
Ref.~\cite{Kessler:2026dbi}
(Fig.~\ref{fig:wz_recon}), the GREA equation of state and the corresponding
dark-energy density follow the reconstructed bins over most of the redshift
range probed, with no freedom beyond $\alpha$, the agreement being closest in
the lower bins where the reconstruction is tightest.

Judged against the standard model on the same benchmark, GREA is
remarkably competitive. It matches the fit of \LCDM\ to within a modest
$\Delta\chi^2$ despite carrying a single $\mathcal{O}(1)$ parameter, and
its Bayesian evidence is comparable throughout, even favouring GREA for the
late-time and CMB-anchored combinations (Table~\ref{tab:grea_best_fit}).
Given that the two models are not nested, and that GREA replaces a
constant tuned by some 120 orders of magnitude below its natural scale with a
thermodynamically motivated, out-of-equilibrium mechanism, achieving parity
with \LCDM\ across CMB, BAO and supernova data is a strong statement in the
model's favour.

Moving to the perturbation level also sharpens the model's confrontation
with the cosmological tensions. When the local distance-ladder calibration
is included, GREA comfortably accommodates a high expansion rate,
$H_0=73.4\pm1.0~\mathrm{km\,s^{-1}\,Mpc^{-1}}$ for PPS+DESI by shifting the
matter-to-acceleration transition to higher redshift. 
This value is largely set by the SH0ES calibration, which acts as a prior on $H_0$, so the PPS+DESI combination does not provide an independent test of the Hubble tension. It does, however, show that GREA remains consistent with the local distance-ladder calibration, fitting the PPS+DESI data marginally better than \LCDM\ ($\Delta\chi^2\simeq-4.7$, Table~\ref{tab:grea_best_fit}).

However, this combination remains in significant tension with the CMB and
therefore does not represent a statistically consistent dataset. For the
consistent combinations of CMB, BAO and Type Ia supernova data, GREA
instead yields values of $H_0$ comparable to those of \LCDM\ and does not
alleviate the Hubble tension.
The $S_8$ sector tells the opposite and more subtle story: the full
Boltzmann computation confirms the enhancement of structure growth foreseen
at the background level, yielding $S_8\simeq0.82$ essentially independent
of the distance calibration (Fig.~\ref{fig:full_triangle}). Whether this
translates into a genuine increase of the $S_8$ tension, however, cannot be
assessed from the present analysis, since the existing weak-lensing
constraints are themselves model dependent and have been derived assuming
\LCDM. A dedicated analysis of the weak-lensing data within the GREA
framework will therefore be required to quantify the level of agreement.
We further show (Appendix~\ref{app:cs2}) that these conclusions are robust
to the one genuinely undetermined ingredient of the analysis, the effective
sound speed of the entropic fluid, which shifts every observable by less
than $1.6\%$, far below current sensitivity.

Several avenues now open naturally. The outstanding theoretical task is the
first-principles derivation of the perturbed entropic sector
$\delta f_{\mu\nu}$, whose only delicate piece is the non-local response of
the light-cone horizon; settling whether it can be neglected would replace
our effective-fluid description by the true GREA perturbations and fix the
low-$\ell$ ISW and lensing response uniquely. Intimately tied to this is
the need to take into account the non-adiabatic perturbations that arise on
the GREA side within linear perturbation theory. Because the entropic force
originates in out-of-equilibrium entropy production, the associated
fluctuations are not, in general, purely adiabatic: the perturbed horizon
thermodynamics can source an intrinsic non-adiabatic pressure that our
effective-fluid description, with its prescribed equation of state and
bracketed sound speed, only approximates. A consistent treatment of these
non-adiabatic modes on the GREA side, derived together with
$\delta f_{\mu\nu}$ rather than imposed by hand, is required to render the
linear theory fully self-consistent and to sharpen the predicted ISW,
lensing and growth signals; this is a demanding task that we leave for
future work. On the observational side, the distinctive growth history of
GREA, its enhanced $\sigma_8$, and its non-monotonic growth index, together
with the ISW and CMB-lensing signals made accessible here for the first
time, will be probed decisively by low-redshift data from DESI, Euclid and
the Vera Rubin Observatory. Equally important will be a dedicated
weak-lensing analysis within the GREA framework, which is required to
establish whether the enhanced growth predicted by the model translates
into a genuine $S_8$ tension with observations. 

The tensor sector, often where modified-gravity models reveal their true
degrees of freedom, is left untouched here but appears benign: because GREA
enters as a source $f_{\mu\nu}$ on the matter side of
Eq.~\eqref{eq:einstein} rather than modifying the geometric side,
it adds no propagating spin-2 mode and no graviton mass, so gravitational
waves keep $c_T=1$ (consistent with GW170817) and the tensor transfer
function is altered only through the background Hubble friction, unlike the
massive-gravity and bigravity theories whose extra spin-2 mode surfaces
precisely in the tensors. The one open question is whether the perturbed
$\delta f_{\mu\nu}$ carries a shear-viscous, transverse-traceless part that
would source tensor anisotropic stress, the tensor-sector counterpart of the
$\delta\zeta=0$ question raised above.

GREA thus remains a
compelling, theoretically grounded alternative to the cosmological
constant, and the Boltzmann-level tools developed here place its sharpest
predictions within reach of the next generation of surveys, which will
ultimately reveal whether our causal horizon drives the present
acceleration and whether the Universe is destined to end in Minkowski
rather than de~Sitter space.

\section*{Acknowledgements}

The authors thank Matteo Martinelli for discussion and useful comments on the draft, and Daniel Kessler for providing the binned reconstruction data of Ref.~\cite{Kessler:2026dbi} shown in
Fig.~\ref{fig:wz_recon}.
SD is funded by MCIN/AEI/10.13039/501100011033 and 
FSE+, reference PRE2021-098098.
DHL is supported by an EPSRC studentship.
EDV is supported by a Royal Society Dorothy Hodgkin Research Fellowship. 
JGB acknowledges support from the Spanish Agencia Estatal de Investigaci\'on under Research Project PID2024-159420NB-C43 [MICINN-FEDER] and the Centro de Excelencia Severo Ochoa Program CEX2020-001007-S at IFT.
This article is based upon work from the COST Action CA21136 - ``Addressing observational tensions in cosmology with systematics and fundamental physics (CosmoVerse)'', supported by COST - ``European Cooperation in Science and Technology''. 
We acknowledge IT Services at The University of Sheffield for the provision of services for High Performance Computing. 

\newpage

\appendix
\section{Entropic acceleration, GREA versus generalized-entropy cosmologies}
\label{sec:grea_vs_genentropy}

Reading the late-time acceleration as a thermodynamic rather than a
vacuum effect defines a broad program, rooted in the observation that
the Einstein equations follow from the Clausius relation applied to a
local Rindler horizon~\citep{Jacobson:1995ab, Padmanabhan:2009vy}. Two
constructions are often grouped under the same entropic label, yet they
make different predictions for the growth of structure, and hence for the
$S_8$ and $H_0$ tensions that frame the CosmoVerse
program~\citep{CosmoVerseNetwork:2025alb, DiValentino:2020vvd}. It is useful to place GREA in relation to the
generalized-entropy models, since the two branches modify opposite sides
of $G_{\mu\nu}=8\pi G\,T_{\mu\nu}$ and treat the second law differently.

The generalized-entropy branch keeps the equilibrium thermodynamic
machinery and deforms the entropy--area law. Applying the Clausius relation
$\delta Q=T\,\mathrm{d}S$ at the apparent horizon
$\tilde r_A=(H^2+k/a^2)^{-1/2}$ with the Hawking temperature
$T=1/(2\pi\tilde r_A)$, and replacing the Bekenstein--Hawking entropy
$S_{\rm BH}=A/4G$ by a one-parameter deformation, carries the modification
into the Friedmann constraint~\citep{Cai:2005ra}. For the Tsallis--Cirto
entropy $S\propto A^{\delta_\text{T}}$~\citep{Tsallis:2012js, Lymperis:2018iuz,
Sheykhi:2018dpn} and the Barrow entropy
$S=(A/A_0)^{1+\delta_\text{T}/2}$~\citep{Barrow:2020tzx, Saridakis:2020zol,
Sheykhi:2021fwh}, the first law gives
\begin{equation}
  -\,\frac{2+\Delta}{2\pi A_0}
   \left(\frac{4\pi}{A_0}\right)^{\!\Delta/2}
   \frac{\mathrm{d}\tilde r_A}{\tilde r_A^{\,3-\Delta}}
  \;=\; \frac{8\pi G \mathrm{d}\rho}{3}\,,
  \label{eq:barrow_diff}
\end{equation}
which integrates to a power-law-deformed first Friedmann equation,
\begin{equation}
  \left(H^2 + \frac{k}{a^2}\right)^{1-\Delta/2}
  \;=\; \frac{8\pi G}{3} \, \rho\ ,
  \label{eq:barrow_friedmann}
\end{equation}
the Tsallis case mapping onto it under $\delta_\text{T}=1+\Delta/2$. The deformation lives
entirely on the geometric side: the matter sector remains adiabatic and
conserved,
\begin{equation}
  \dot\rho + 3H(\rho + p) = 0\,,
  \label{eq:genentropy_continuity}
\end{equation}
and the departure from \LCDM\ is bookkept as an effective component,
$\rho_{\rm DE}(H,\dot H)$, assembled from curvature- and rate-dependent
terms rather than a genuine flux. Such models have been used as
tension-easing scenarios, the Tsallis case in particular being proposed to relax both the $H_0$
and $S_8$ tensions~\citep{Basilakos:2023kvk}, though the exponents
$\delta_\text{T}$ and $\Delta$ are fitted to the data rather than predicted, and the
acceleration is achieved at the expense of additional free parameters.

GREA acts on the matter side instead. It leaves the Bekenstein--Hawking
area law intact and drops the equilibrium assumption, making the expansion
irreversible through the covariant non-equilibrium thermodynamics of
\citep{Garcia-Bellido:2021idr, Espinosa-Portales:2021cac}. The
entropy growth of the homogeneous causal horizon enters the matter side of
the field equations as the entropic-force tensor $f_{\mu\nu}$ of
Eq.~\eqref{eq:einstein}, a bulk-viscous term with a negative effective
pressure $p_S$; equivalently, the source of curvature is the Helmholtz free
energy $\mathcal{F}=U-TS$ rather than the internal energy
alone~\citep{Garcia-Bellido:2024qau}. Because entropy production does not
vanish, time-reversal invariance is broken and the continuity equation
contains the source term of Eq.~\eqref{eq:continuity}, in contrast to the
conserved evolution of Eq.~\eqref{eq:genentropy_continuity}. At the
background level, this reduces to the closed system of
Sec.~\ref{sec:framework}: the expansion history of
Eq.~\eqref{eq:background}, set by the single $\mathcal{O}(1)$ parameter
through Eq.~\eqref{eq:alpha}, with the effective equation of state of
Eq.~\eqref{eq:weff} and its transient phantom crossing that does not
violate the null energy condition~\citep{Garcia-Bellido:2024qau,
Calderon:2025dhj}. The phenomenology is predicted rather than fitted:
$\alpha\sim1$ ties the size of the causal horizon today to the curvature
scale, and every background observable inherits its dependence on this one
parameter.

The cleanest observational discriminator between the two branches is the
growth of matter perturbations. In GREA the linear growth is modified by
neither dark-energy clustering nor an effective fifth force, but follows
entirely from the altered expansion history through the Hubble-friction
term of Eq.~\eqref{eq:growth}; the combination $\fsig(z)$ therefore tracks
\LCDM\ closely, and the discriminating information is carried by the growth
index $\gamma(z)\equiv\ln f(z)/\ln\Omega_{\rm m}(z)$. The companion
background analysis~\citep{Garcia-Bellido:2024qau,Calderon:2025dhj} finds a
present-day value $\gamma\simeq0.55$, essentially indistinguishable from
\LCDM, but a qualitatively different redshift evolution,
$\mathrm{d}\gamma/\mathrm{d}z>0$, in sharp contrast with the monotonically
{\it decreasing} $\gamma(z)$ of \LCDM\ and most quintessence and
modified-gravity models~\citep{Polarski:2016ieb,Calderon:2019jem,Calderon:2019vog}.
The generalized-entropy cosmologies instead alter the growth through their
deformed Friedmann constraint acting on an otherwise conserved matter
sector, and can be tuned to {\it suppress} $\sigma_8$ and relax the $S_8$
tension~\citep{Basilakos:2023kvk}; GREA does the opposite, enhancing the
growth at fixed $A_s$ (Sec.~\ref{sec:perturbations}) and predicting the
rising $\gamma(z)$ described above. The sign of the growth-index slope is
thus a sharp, falsifiable signature that separates the two entropic
scenarios from each other and from \LCDM, and forthcoming
redshift-space-distortion and weak-lensing measurements from DESI, Euclid
and the Vera Rubin Observatory~\citep{Amendola:2016saw} will measure it
with the required precision.

The two pictures therefore differ in where the new physics resides.
Generalized-entropy cosmologies deform the static equation of state of the
horizon and modify the Friedmann constraint while leaving conservation
intact; GREA keeps the area law but changes the thermodynamic regime of
the bulk, adding an irreversible flux that acts as a negative-pressure
bulk-viscous source in the continuity and acceleration equations. The
modification is geometric in the first case and enters the stress-energy
tensor in the second, and only in GREA is the second-law bookkeeping
non-trivial: the generalized-entropy dark-energy term relabels
curvature- and expansion-rate-dependent geometry, whereas $f_{\mu\nu}$ is
a genuine entropy-producing current. This is why the appropriate
Boltzmann-code realization of GREA is an effective-fluid dark-energy
sector with its own perturbations, rather than the modified-friction
recasting that suffices for the generalized-entropy family, and why a
Horndeski-type scalar-tensor implementation would not be appropriate.

\section{Sensitivity to the dark-energy sound speed}
\label{app:cs2}

The parametrized post-Friedmann (PPF) treatment of the GREA effective fluid
requires an effective sound speed $c_s^2$, which sets the scale below which the
fluid stops clustering and follows a smooth sub-sound-horizon evolution. The
GREA closure for the perturbed stress tensor $\delta f_{\mu\nu}$ does not fix
this quantity, so throughout the main analysis we adopt the standard value
$c_s^2=1$. Since $c_s^2$ enters only through the clustering transition and
leaves the background, and therefore every BAO and supernova distance,
unchanged, its influence is restricted to two observables: the low-$\ell$
temperature spectrum through the late-time integrated Sachs--Wolfe (ISW)
effect and the lensing amplitude. We quantify that influence here and show
that it lies far below the sensitivity of the data used in this work.

Holding all cosmological parameters at the CMB-SPA+PP+DESI best fit, we vary
$c_s^2$ over a dense grid covering the physically allowed range
$c_s^2\in[0.01,1]$ and compute the fractional residual of each observable
relative to the $c_s^2=1$ baseline. In the PPF scheme, $c_s^2$ is not a
rest-frame pressure closure but the scale governing the clustering-to-smooth
transition, so this range brackets the full span from a fluid that clusters
down to small scales ($c_s^2\to0$) to one that is smooth above the sound
horizon ($c_s^2=1$).

\begin{figure}[t]
  \centering
  \includegraphics[width=\columnwidth]{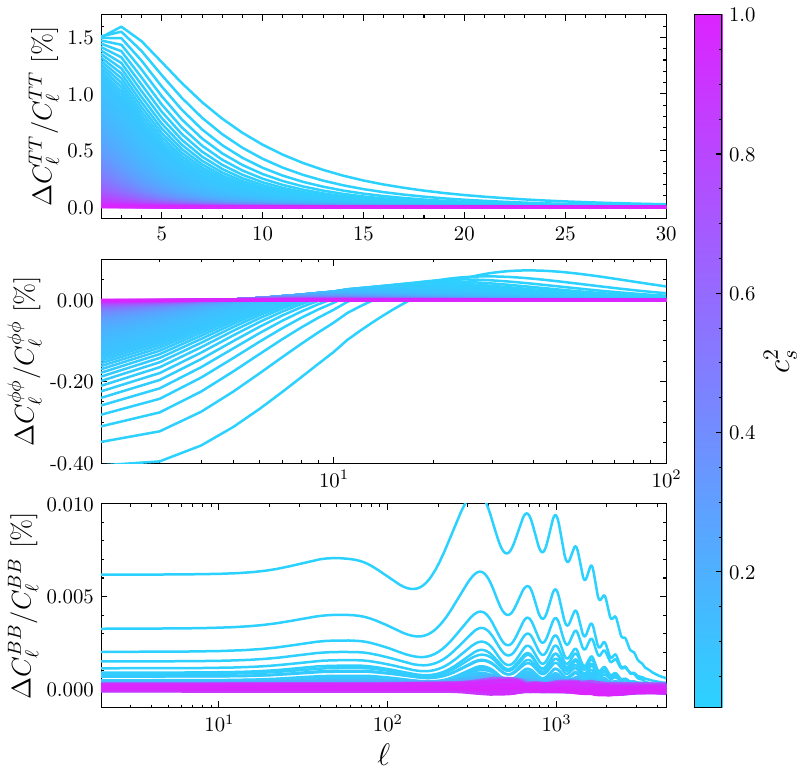}
  \caption{Fractional residuals of the CMB spectra as the dark-energy sound speed
  is varied over $c_s^2\in[0.01,1]$ at the fixed CMB-SPA+PP+DESI best-fit cosmology,
  relative to the $c_s^2=1$ (smooth) baseline; the line color runs from strongly
  clustering (cyan) to smooth (violet). \textit{Top}: low-$\ell$ temperature
  spectrum, where the late-time ISW response is largest, shown against the
  $\pm1\sigma$ cosmic-variance floor (shaded). \textit{Middle}: lensing potential
  $C_\ell^{\phi\phi}$; every
  residual remains well within the cosmic variance and below the lensing
  uncertainties of current experiments. \textit{Lower}: lensing-induced
  $C_\ell^{BB}$.}
  \label{fig:cs2_main}
\end{figure}

Figure~\ref{fig:cs2_main} collects the CMB response. The low-$\ell$ temperature
residual reaches at most $1.6\%$ near $\ell\simeq3$ and falls to a fraction of a
percent by $\ell\simeq15$. The lensing potential $C_\ell^{\phi\phi}$ and the
lensing-induced $C_\ell^{BB}$ remain below $0.6\%$ and $0.2\%$, respectively,
and are negligible across the multipole range where CMB lensing carries
statistical weight. All the curves lie well below the cosmic variance, which
reaches $\pm76\%$ at $\ell=2$; the largest residual anywhere in the sweep is
roughly forty times smaller than this floor at the same multipoles, and lies
well below the per-band lensing uncertainties of Planck, ACT, and SPT. Cross-correlating
the low-$\ell$ temperature with galaxy surveys, the standard route to beating
the ISW cosmic-variance floor, does not change this conclusion: even a
Stage~IV galaxy$\times$TT measurement detects the full ISW signal at only
$\lesssim10\sigma$, far short of the sensitivity needed to resolve a
percent-level modulation of it, so the $c_s^2$ ambiguity remains observationally
irrelevant for the foreseeable future. This should not be confused with the
much larger $\alpha$-driven growth and ISW signal of Figs.~\ref{fig:power_spectrum_1}
and~\ref{fig:power_spectrum_2}, a tens-of-percent effect that Stage~IV
surveys will probe decisively.

The sign of the temperature response is opposite to the naive quintessence
expectation, in which dark-energy clustering suppresses the ISW contribution and
lowers the low-$\ell$ power. At the CMB-SPA+PP+DESI best fit, the GREA equation
of state is phantom, $w<-1$, over the redshift range where the ISW signal is
generated. This reverses the sign of $(\rho+p)$ and hence of the coupling
between dark-energy clustering and the evolving gravitational potential,
mirroring the inversion of the background-level intuition already seen for
structure growth in Sec.~\ref{sec:perturbations}.

Taken together, the un-derived sound speed of the GREA effective fluid changes
every observable entering this analysis by less than $1.6\%$, more than an
order of magnitude below the sensitivity of the current data. The baseline
$c_s^2=1$ constraints are therefore robust to this choice. A parameter-level
check with $c_s^2\to0$ would leave the data vector unchanged relative to its
errors and so cannot shift the posteriors, and we do not pursue it here.

\section{Additional plots}\label{app:additional}

\begin{figure*}[ht]
\centering
\includegraphics[width=\textwidth]{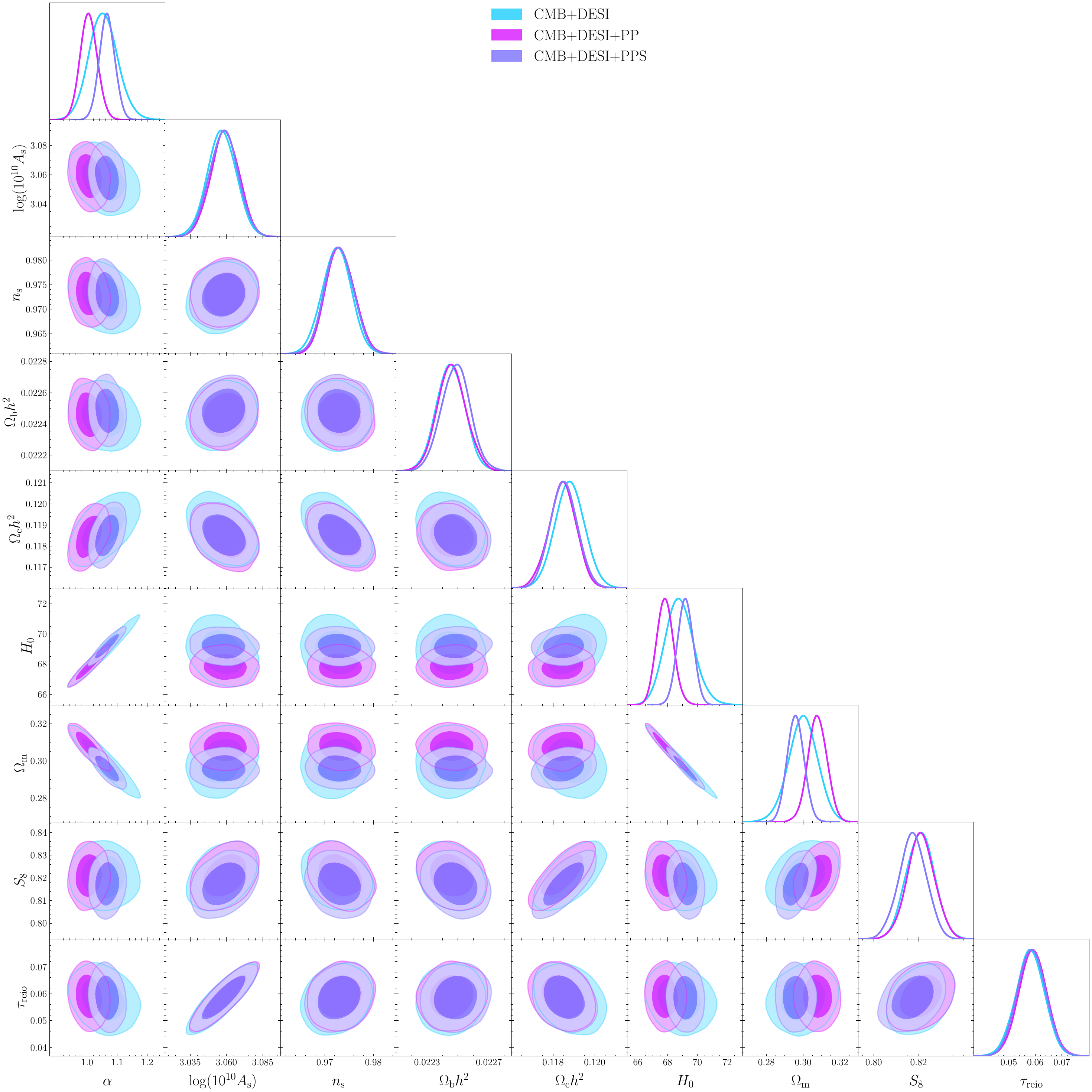}
\caption{Marginalized one- and two-dimensional posterior distributions for all sampled and key derived parameters in the GREA model. The inclusion of late-time data tightens the constraints on all parameters, while the SH0ES-calibrated Pantheon+ sample shifts $H_0$ towards higher values.}
\label{fig:full_triangle}
\end{figure*}

\begin{figure*}[ht]
\centering
\includegraphics[width=\textwidth]{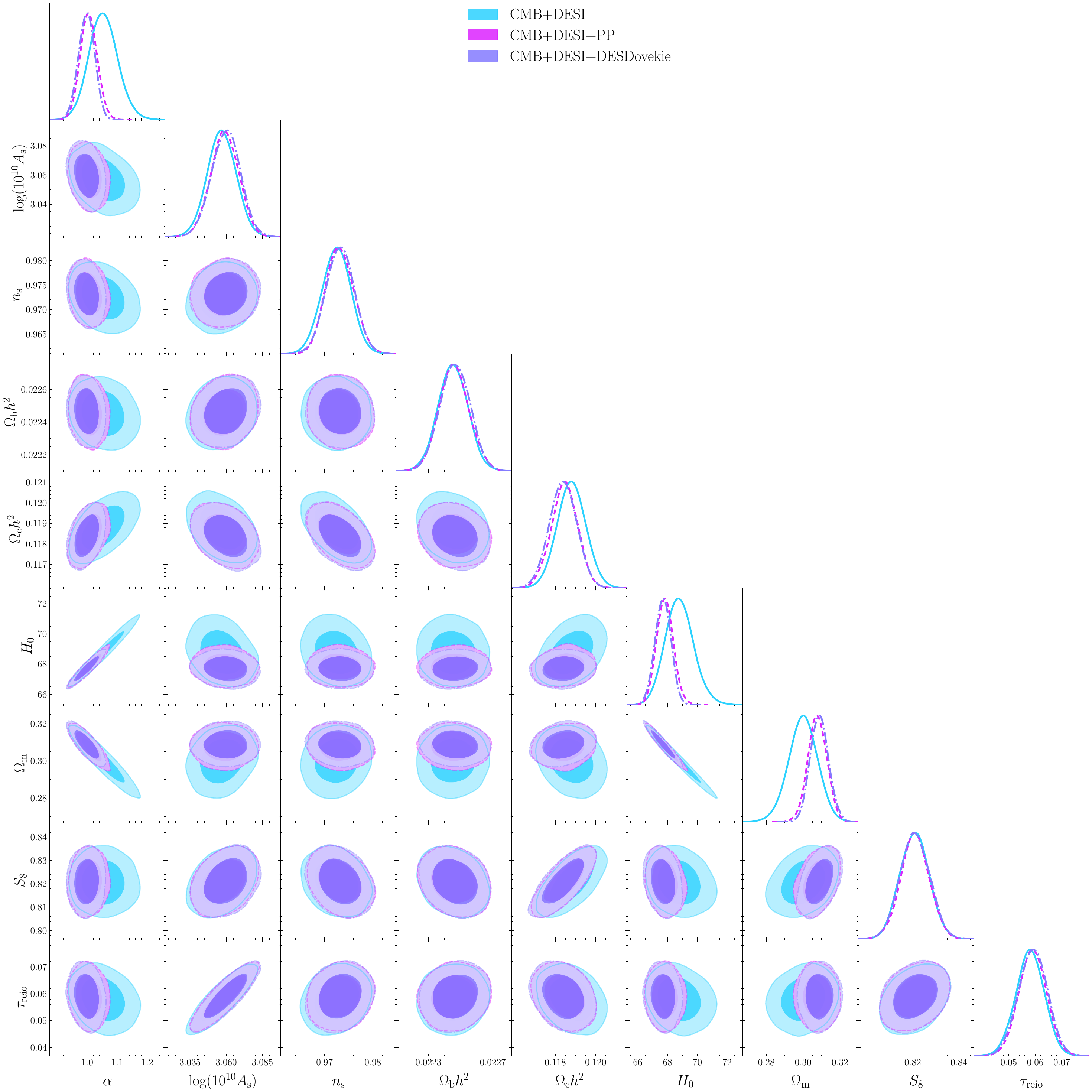}
\caption{Marginalized one- and two-dimensional posterior distributions for all sampled and key derived parameters in the GREA model. Comparison between the Pantheon+ and recalibrated DES Dovekie Type Ia supernova compilations.}
\label{fig:full_triangle_Sne}
\end{figure*}

\begin{figure*}[ht]
\centering
\includegraphics[width=\textwidth]{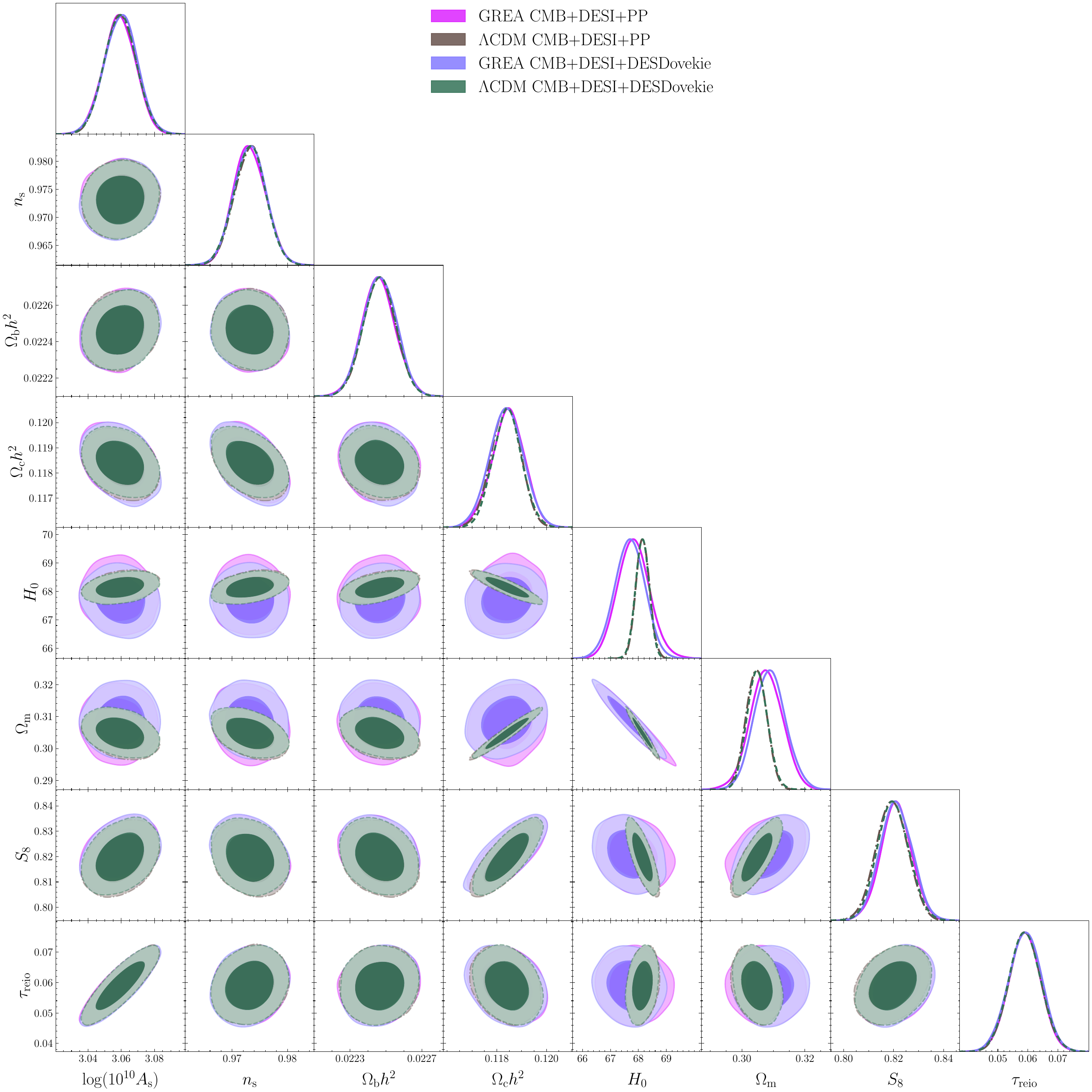}
\caption{Marginalized one- and two-dimensional posterior distributions for all sampled and key derived parameters in the GREA model. Comparison between different Type Ia supernova compilations (Pantheon+ and DES Dovekie) and the $\Lambda$CDM model.}
\label{fig:full_triangle_LCDM}
\end{figure*}

For completeness we collect here the full marginalized posterior distributions
for all sampled and derived parameters, of which the headline subspaces shown in
Sec.~\ref{sec:results} are projections. Figure~\ref{fig:full_triangle} compares
the three CMB-anchored combinations, Fig.~\ref{fig:full_triangle_Sne} the two
supernova compilations, and Fig.~\ref{fig:full_triangle_LCDM} places the GREA
posteriors alongside their $\Lambda$CDM counterparts for the same data.

Two features are worth noting beyond what the headline parameters already show.
The GREA coupling is essentially uncorrelated with the early-time parameters
$\omega_b$, $\omega_{cdm}$, $n_s$ and $\log(10^{10}A_s)$, its degeneracy
directions being confined to the late-time sector through $\Omega_m$ and $H_0$,
which is why the inferred early-time cosmology is left unchanged with respect to
$\Lambda$CDM. The broadening of the contours in GREA, roughly a factor of two on
$H_0$ and $\Omega_m$, is likewise confined to the parameters that enter this
degeneracy, while the remaining posteriors are close to indistinguishable from
the $\Lambda$CDM ones.

\addtocontents{toc}{\setcounter{tocdepth}{-10}}
\newpage
\quad
\phantomsection

\bibliographystyle{apsrev4-1}

\bibliography{biblio}

\end{document}